\documentclass[aps,prb,twocolumn,superscriptaddress,showpacs]{revtex4}

\usepackage{graphicx}
\usepackage{amsmath}

\begin{document}

\title{Quantum and semiclassical study of magnetic anti-dots}

\author{Bence Kocsis}
\affiliation{Department of Physics of Complex Systems, E{\"o}tv{\"o}s
University, H-1117 Budapest, P\'azm\'any P{\'e}ter s{\'e}t\'any 1/A,
Hungary}
\author{Gergely Palla}
\affiliation{Biological Physics Research Group of HAS, E{\"o}tv{\"o}s
University, H-1117 Budapest, P\'azm\'any P{\'e}ter s{\'e}t\'any 1/A,
Hungary}

\author{J\'ozsef Cserti}
\email{cserti@galahad.elte.hu}
\affiliation{Department of Physics of Complex Systems, E{\"o}tv{\"o}s
University, H-1117 Budapest, P\'azm\'any P{\'e}ter s{\'e}t\'any 1/A,
Hungary}


\begin{abstract}

We study the energy level structure of
two-dimensional charged particles in inhomogeneous magnetic fields.
In particular, for magnetic anti-dots the magnetic field
is zero inside the dot and constant outside.
Such a device can be fabricated with present-day technology.
We present detailed semiclassical studies of such magnetic anti-dot
systems and provide a comparison with exact quantum calculations.
In the semiclassical approach we apply the Berry-Tabor formula for the
density of states and the Borh-Sommerfeld quantization rules.
In both cases we found good agreement with the exact spectrum in
the weak magnetic field limit.
The energy spectrum for a given missing flux quantum is classified
in six possible classes of orbits and summarized
in a so-called phase diagram.
We also investigate the current flow patterns of different quantum
states and show the clear correspondence with classical trajectories.

\end{abstract}


\maketitle

\section{Introduction}\label{intro:ref}

In the past decade, the study of systems of two-dimensional 
electron gas (2DEG) in semiconductors\cite{Houten}
has been extended by
the application of spatially inhomogeneous magnetic fields.
The inhomogeneity of the magnetic field can be realized
experimentally either by varying the topography of the
electron gas~\cite{Foden:cikk,Leadbeater-1:cikk}, by
using ferromagnetic materials~\cite{Leadbeater-2:cikk,Krishnan:cikk,Nogaret-1:cikk,Nogaret-2:cikk,Nogaret-4:cikk,Nogaret-3:cikk},
depositing a superconductor on top of the 2DEG~\cite{Smith:cikk,Geim:cikk}.
Numerous theoretical works also show an increasing interest in the study
of electron motion in inhomogeneous magnetic field
(see, e.g., Refs.~\onlinecite{Muller-1:cikk,Khveshchenko:cikk,Gotz:cikk,Peeters-1:cikk,Calvo:cikk,Peteers-2:cikk,Schmelcher:cikk,Matulis:cikk,Nielsen:cikk,Peteers-3:cikk,Sim:cikk,Urbach:thesis,Kim-1:cikk,Michele_Boese:cikk,Kim-2:cikk,Kim-3:cikk,Peteers-4:cikk,Peteers-5:cikk,Klaus_spin-orbit:cikk,IMB:cikk}).

In the experimental works mentioned above, for GaAs heterostructures,
on the one hand, the electron dynamics is confined to two-dimensions.
On the other hand, the coherence length and the mean free path of the
electron can be much larger than the
size of the system, while the Fermi wavelength
is comparable to the size of the 2DEG.
Moreover, the electron system can be described to a good approximation as
a free electron gas with an effective mass\cite{Houten}.
Therefore, the quantum mechanical treatment of these systems
is of some physical interest.

\begin{figure}[hbt]
\includegraphics[scale=0.4]{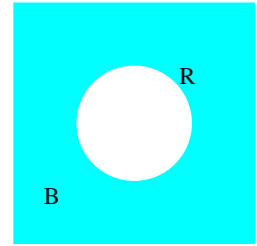}
\caption{The two-dimensional electron gas in an inhomogeneous magnetic 
field.
The magnetic field is zero inside the circle of radius $R$ and
constant outside the circle.
\label{geo_Bantidot:fig}}
\end{figure}

In this paper, as an example, we consider the energy levels of a 
two-dimensional non-interacting electron gas in a
magnetic field that is zero inside a circular region
and constant outside.
This system (shown in Fig.~\ref{geo_Bantidot:fig})
will be called a magnetic anti-dot; it was first studied by
Solimany and Kramer~\cite{Solimany}. Solving the Schr{\"o}dinger
equation it was shown that there are bound states.
Introducing an effective angular momentum,
the Schr{\"o}dinger equation of the particle in symmetric gauge can
be mapped to the Landau model.
This effective angular momentum is a sum of the angular momentum in
a uniform magnetic field and the flux (in units of
the flux quantum) missing from the uniform field.
Recently, Sim et al.~\cite{Sim:cikk} have renewed the study of
this system and pointed out the crucial role of the magnetic edge states
in the magnetoconductance.
The classical counterparts of these states correspond to trajectories of
the charged particle that consist of straight segments inside the
non-magnetic region and arcs outside.

Although it is difficult to measure directly the density of states of
a quantum system, it affects many observable quantities such as
the magnetoconductance, the magnetization or the susceptibility.
In the interpretation of the experimental results the semiclassical
approximation proved to be a useful tool.
Several semiclassical
approaches~\cite{WKB:cikk,Keller:cikkek,Berry-Tabor:cikk,Balian_Bloch:cikkek,Gutzwiller:cikkek,Strutinsky:cikkek,Creagh_Littlejohn:cikk}
are known in the literature and an excellent overview of
the subject can be found in the textbook by Brack and Bhaduri\cite{Brack:book}.
Different semiclassical theories for magnetic systems have
successfully been applied, for example,
in works~\cite{Blaschke:cikk,Klaus_H,Klama,Klama-2}.
For integrable systems Berry and Tabor~\cite{Berry-Tabor:cikk}
have shown that the oscillating part of the density of states
can always be expressed in terms of classical periodic orbits.
This formula is commonly called the Berry-Tabor trace formula.

One of our aims in this paper is to apply, for the first time,
the Berry-Tabor trace formula for a magnetic anti-dot.
To illustrate the power of the method,
we also calculate the exact eigenvalues of the single particle
Schr{\"o}dinger equation and find a very good agreement between the two
results.
We should mention here that the statement by Sim et al.~\cite{Sim:cikk}
on the relation between the quantum states and the
corresponding periodic orbits is somewhat
misleading.
Their condition for a given periodic orbit is not necessarily satisfied
at the value of the corresponding exact energy level as they claimed 
However, including more and more periodic orbits with the proper weights
in the trace formula the sum converges to the correct quantum density of
states. In practice only a few of the shortest orbits are
enough to get a rough estimate of the positions of the
exact energy levels.

The power of the semiclassical approach can also be demonstrated by
applying the Bohr-Sommerfeld approximation.
We shall show that the energy levels obtained from the Bohr-Sommerfeld
quantization rules also agree very well with the numerically exact levels
even for the lowest eigenstates.
Note that in this semiclassical treatment the quantization should be
applied to the classical motion on a two-dimensional torus
parametrized by the action variables and their canonically conjugate
angle variables (for details see, eg, Ref.~\onlinecite{Brack:book}).

The classical orbits can be classified by their cyclotron radius $\varrho$
and their guiding center $c$ (distance of the center of the orbits
from the origin).
In the quantum mechanical treatment one can calculate the average of
operators defining the cyclotron radius and the guiding center.
For circular magnetic billiards, Lent~\cite{Lent:cikk} has derived
approximate expressions for these averages for a given quantum state.
Following Ref.~\onlinecite{Lent:cikk}, one may derive the corresponding
relations for magnetic anti-dot systems.
Thus, these relations are the basis for classifying the different quantum
states in terms of classical orbits for our system.
We will show that the quantum states can be described by six different
types of classical orbits.
In addition, this classification enables us to draw
a so-called `phase diagram' which shows a clear one-to-one correspondence
between classical orbits and quantum states.

To complete our semiclassical study, we finally present results for
the probability current density calculated from quantum calculations.
We shall argue that the current flow patterns can be understood
qualitatively from the corresponding classical trajectories.
Recently, Halperin~\cite{Halperin:cikk} has been shown that
the total current (the integral of the current density along the
radial direction) can be related to the dispersion of the energy levels
(their angular momentum dependence).
As it will be shown this general relation works in our magnetic
anti-dot system, too.

Regarding the numerical calculations, we should mention that
the semiclassical approach presented in this paper proves to
be a very effective method.
Moreover, it provides a better understanding of the nature of the
quantum system.
Our semiclassical method applied to  magnetic anti-dot systems
may be an important tool to understand the
role of the magnetic edge states in the density of states or
the magnetization (both are experimentally accessible physical qauntities).
We believe that our semiclassical analysis can be extended
to other types of inhomogeneous magnetic fields
such as studied, eg, in
Refs.~\onlinecite{Nogaret-1:cikk,Nogaret-2:cikk,Nogaret-4:cikk,Nogaret-3:cikk}
as well as non-circular dot systems.

The rest of the text is organized as follows. In Sec.\ \ref{quantum:seq}
the exact quantization condition (secular equation) is derived from
the matching conditions of the wave functions at the boundary of the
magnetic and non-magnetic regions.
In Sec.~\ref{Berry-Tabor:ref} the semiclassical approximation is
presented including the description of the classical motion of the
particle in Subsec.~\ref{class_dyn:sec},
the characterization of the possible periodic orbits
in Subsec.~\ref{periodic_orbit:sec}, some numerical results
in Subsec.~\ref{Berry_Tabor_results:sec}, and the
phase diagram in Subsec.~\ref{phase_diag:sec}.
The current flow patterns of the system are discussed
in Sec.~\ref{current:sec}.
Finally, the conclusions are given in Sec.~\ref{conclusion:sec}.

\section{Quantum calculation}\label{quantum:seq}

In this section, we present the quantum mechanical treatment of the
magnetic anti-dot.
The magnetic field with a constant $B$ outside
a circle of radius $R$ is assumed to be
perpendicular to the plane of the 2DEG.
The Hamiltonian of the electron of mass $M$ and charge $e$ is given by
\begin{equation}
H = \left\{ \begin{array}{ll}
\frac{{\bf p}^2}{2M},
& {\rm if}\,\,\,\,  r < R,   \\[2ex]
\frac{{\left({\bf p} - e{\bf A}\right)}^2}{2M},
& {\rm if} \,\,\,\, r > R ,
\end{array}
\right.
\label{H_op-start:eq}
\end{equation}
where ${\bf p}$ is the canonically conjugate momentum,
and  the vector potential in polar coordinates $(r,\varphi)$ and
symmetric gauge is  given by~\cite{Solimany}
\begin{eqnarray}
{\bf A} &=& A_\varphi(r,\varphi) \, \hat{{\bf e}}_\varphi ,
\,\,\, \text{where}
\label{vector-pt:eq} \\[1ex]
A_\varphi(r,\varphi) &=& B\, \frac{r^2 -R^2}{2r}\,\Theta(r-R), \nonumber
\end{eqnarray}
and $\hat{{\bf e}}_\varphi$ is the unit vector in the $\varphi$ direction.
Here $\Theta(x)$ is the Heaviside step function.

The energy levels of the system are the eigenvalues
$E$ of the Schr{\"o}dinger equation:
\begin{equation}
\hat{H} \Psi (r,\varphi) = E \Psi (r,\varphi).
\label{Sch_start:eq}
\end{equation}

Rotational symmetry of the system implies a separation ansatz
for the wave function as a product of radial and angular parts.
We choose for the angular part the appropriate angular momentum 
eigenfunctions $e^{im\varphi}$ with quantum number $m$ 
(here $m$ is an integer).
Thus the wave function for a given $m$ is separated
as $\Psi(r,\varphi) = f_m(r)
e^{im\varphi}$, where the radial wave functions
$f_m(r)$ satisfy a one-dimensional
Schr{\"o}dinger equation in the normal region:
\begin{subequations}
\begin{equation}
\hat{h}_m (\tau) f_m(\tau) =  \varepsilon f_m(\tau),
\label{rad_Sch:eq}
\end{equation}
in which the radial Hamiltonian  takes the form
\begin{equation}
\hat{h}_m (\tau) = -\frac{\partial^2}{\partial \tau^2}
-\frac{1}{\tau}\frac{\partial}{\partial \tau}
+V_m(\tau).
\label{H_rad_op:eq}
\end{equation}
Here  we introduce the dimensionless variable $\tau = r/l$, where
$l=\sqrt{\hbar /|eB|}$ is the magnetic length,
$\omega_c = |eB|/M$ is the cyclotron frequency,
$\varepsilon = 2E/(\hbar \omega_c)$ is the dimensionless energy,
and the radial potential is given by
\begin{eqnarray}
V_m(\tau)  &=& \left\{ \begin{array}{ll}
\frac{m^2}{\tau^2},  & {\rm if} \,\,\,\,  r < R,  \\[2ex]
\frac{{\left(\frac{\tau^2}{2}
-m_{\rm eff}\right)}^2}{\tau^2}, & {\rm if} \,\,\,\,  r > R,
\end{array}
\right.
\label{Vm-pot:eq} \\
m_{\rm eff} &=& s + m\, \rm{sgn}(eB) ,
\label{m_eff:def}
\end{eqnarray}
\end{subequations}
where $s = R^2/(2 l^2) = \Phi/\Phi_0 $ is the magnetic flux
$\Phi = B R^2 \pi $ (in units of the magnetic flux quantum
$\Phi_0 = h/\left|e\right|$)
missing  inside the circle of radius $R$.
The function $\rm{sgn} \{\cdot\}$ stands for the sign function.
In the numerical results presented in this paper, we always assume
that the particle is an electron moving in a magnetic field
along the positive $z$-axis, i.e., $\rm{sgn}(eB) = -1$.
However, our theoretical results are not restricted in such a way.

Introducing the new variable $\xi=\tau^2/2$ and transforming
the wave functions in the magnetic region ($r>R$) as
\begin{equation}
f_m(\tau) = \xi^{\mid m_{\rm eff}\mid/2} \, e^{-\xi/2}\, \chi_m(\xi),
\end{equation}
Eq.~(\ref{rad_Sch:eq}) results in
a Kummer differential equation~\cite{Abramowitz}
\begin{eqnarray}
\xi \frac{d^2 \chi_m}{d \xi^2}
+\left(1 + \mid m_{\rm eff}\mid -\xi \right)\frac{d \chi_m}{d\xi}
&&  \nonumber \\
- \frac{1+\mid m_{\rm eff}\mid -m_{\rm eff} - \varepsilon}{2}
\chi_m &=& 0.
\end{eqnarray}
Thus the ansatz for the radial wave function in the magnetic region
can finally be written as
\begin{eqnarray}
\lefteqn{f_m(\tau) = \xi^{\mid m_{\rm eff}\mid/2} \, e^{-\xi/2} }
&& \nonumber \\
&& \times \,
U\left( \frac{1+\mid m_{\rm eff}\mid -m_{\rm eff} - \varepsilon,}{2},
1 + \mid m_{\rm eff} \mid,\xi \right),
\end{eqnarray}
where $U$ is the confluent hypergeometric function~\cite{Abramowitz}.
Note that the function $U$ tends to zero as $r \rightarrow \infty$.

It is easy to show that the radial wave function inside the circle of
radius $R$ (where the magnetic field is zero) satisfies
the Bessel differential equation\cite{Abramowitz}.
Thus the radial wave function is given by
\begin{equation}
g_m(\tau) = J_m(\sqrt{\epsilon}\, \tau),
\end{equation}
where $J_m(x)$ is the Bessel function of order $m$.

Matching the radial wave functions inside and outside the circle gives
a secular equation whose solutions are the eigenvalues of the system.
The matching conditions at $r = R$ yield
\begin{equation}
\frac{d}{d \tau} \, \ln g_m (\tau)\,
\rule[-1.6ex]{.2pt}{4ex}
\;\raisebox{-1.5ex}{$\scriptstyle \tau=R/l$}
= \frac{d}{d \tau} \, \ln f_m (\tau)\,
\rule[-1.6ex]{.2pt}{4ex}
\;\raisebox{-1.5ex}{$\scriptstyle \tau=R/l$}.
\label{secular:eq}
\end{equation}
For a given $m$ this secular equation depends only on the dimensionless
missing flux $s$.

\section{Semiclassical approximation: the Berry-Tabor approach}
\label{Berry-Tabor:ref}

We now turn to the semiclassical treatment of the system.
Generally, in $d$ dimensions, a  system is integrable if there are
$d$ independent constants of the motion. Usually this is the
result of the separability of the Hamiltonian:
In a suitably chosen coordinate system the Hamiltonian depends
only on separate functions $\phi_i(q_i,p_i)$ of the coordinates
and the conjugate momenta. This means that the dynamics can be
viewed as a collection of independent one dimensional dynamical
systems. The function $\phi(q_i,p_i)$ plays the role of the
Hamiltonian in each subsystem. The one dimensional
semiclassical quantization procedure can be carried out in each
subsystem separately
\begin{equation}
I_i=\frac{1}{2\pi}\oint 
p_idq_i=\hbar\left(n_i+\frac{\nu_i}{4}\right),\;\;\;n_i=0,1,2,...,
\label{action}
\end{equation}
where $I_i$ is the action variable and $\nu_i$ is the Maslov index
(for details see, eg, Ref.~\onlinecite{Brack:book}).
The Maslov index is the sum of the Maslov indices of the turning
points of the classical motion. Smooth or ``soft'' classical turning points
(zeros of $p_i(q_i)$) contribute $+1$ to the Maslov index, while
``hard'' classical turning points (infinite potential walls)
contribute $+2$.
Equation~(\ref{action}), the Bohr-Sommerfeld quantization condition, is 
widely used to approximate the energy levels of classically integrable
systems.

Alternatively, from Eq.~(\ref{action}) a semiclassical trace formula
known as the Berry-Tabor formula~\cite{Berry-Tabor:cikk}
can be derived for the oscillating part of the density of states.
For two-dimensional systems, this formula can be written as
\begin{eqnarray}
d(E)&=&d_0(E) \nonumber \\ & &
+ \sum_{p}\sum_{j=1}^{+\infty}
\frac{\cos\left({\frac{jS_p(E)}{\hbar}}-\frac{\pi}{2}j\nu_p+\frac{\pi}{4}
\right)}{\pi \hbar^{3/2}\sqrt{
\frac{j (n_{2,p})^3 }{T_p^2}\, \frac{\partial^2 g}{\partial I_1^2}}},
\label{BerryTabor2d}
\end{eqnarray}
(for the detailed derivation of this expression see
Appendix~\ref{Berry_Tabor:app}).
Here $d_0(E)$ is the average density of states.
The $p$-summation runs over the primitive periodic orbits of the system, 
the $j$-summation runs over their repetitions; $S_p,T_p$ and
$\nu_p$ denote the classical action, the time and the Maslov-index of orbit
$p$, respectively;
$n_{2,p}$ is the number of cycles in the motion
projected to the action variable $I_2$ under one cycle of the orbit; and
$I_1=g(I_2,E)$ denotes the action variable $I_1$ as a function of the
energy and $I_2$.

\subsection{Classical dynamics of the system}
\label{class_dyn:sec}

It is easy to show that the classical Hamiltonian
in polar coordinates $(r,\varphi)$, inside and outside
the non-magnetic region is:
\begin{equation}
H = \frac{p_r^2}{2M} + V(r),
\label{class_H:eq}
\end{equation}
where $p_r$ and $p_\varphi$ are the canonically conjugate momenta,
and the radial potential $V(r)= \frac{\hbar \omega_c}{2}\, V_m(\tau)$
is the same as in (\ref{Vm-pot:eq})
with the following replacements:
\begin{subequations}
\begin{eqnarray}
m & = & \frac{p_\varphi}{\hbar}, \\
m_{\rm eff} & = & s+ \frac{p_\varphi}{\hbar} \, \rm{sgn}(eB).
\end{eqnarray}
\label{replace_m:eq}
\end{subequations}
Note that here $m$ and $m_{\rm eff}$ are continuous classical variables.
As we shall see below in the semiclassical approximation,
the canonical momentum $p_\varphi$ is quantized according to
the Bohr-Sommerfeld quantization rules (\ref{action}).

Since the Hamiltonian does not depend explicitly on $\varphi$ (the system
is rotationally invariant), the conjugate momentum $p_\varphi$ is a constant
of motion.
Thus the angular action variable becomes
\begin{equation}
I_\varphi = \frac{1}{2\pi} \, \oint p_\varphi\,  d\varphi = p_\varphi.
\label{angular_action:eq}
\end{equation}
The conjugate momentum inside the anti-dot is in fact the angular
momentum. Outside the anti-dot there is an additional term due to the
non-zero vector potential. 
From $\dot{\varphi}= \partial H / \partial p_\varphi$ one finds
\begin{equation}
p_\varphi = \left\{ \begin{array}{lr}
Mr^2 \dot{\varphi}, &  {\rm if} \,\,\,\,  r < R,   \\[2ex]
Mr^2 \dot{\varphi} + eB\, \frac{r^2-R^2}{2}, &  {\rm if} \,\,\,\,  r > R.
\end{array}
\right.
\label{pfi_phi_dot:eq}
\end{equation}

We now choose $I_1 \equiv I_r$ and $I_2 \equiv I_\varphi$
in Eq.~(\ref{action}).
To calculate the radial action variable $I_r = g(I_\varphi,E)$
one needs to perform the integral of $p_r=\sqrt{E-V(r)}$
between the classical turning points of the radial potential.
For a given $E$ these turning points can be obtained from $V (r)=E$.
Using the same dimensionless variables as in Sec.~\ref{quantum:seq},
we have one turning point for the potential inside the non-magnetic circle:
\begin{subequations}
\begin{equation}
\tau_0^{\rm in} = \frac{m}{\sqrt{\varepsilon}},
\label{turnpoints-in:eq}
\end{equation}
and for the potential valid outside the circle there are two turning points:
\begin{eqnarray}
\lefteqn{ \tau_{1,2}^{\rm out}  =
\sqrt{2(\varepsilon+m_{\rm eff}) \mp 2\sqrt{\varepsilon
(\varepsilon + 2m_{\rm eff})}}
}
&& \nonumber \\
&=& \!\! \sqrt{\varepsilon+m_{\rm eff}+\left|m_{\rm eff}\right|}
\mp\sqrt{\varepsilon+m_{\rm eff}-\left|m_{\rm eff}\right|} ,
\label{turnpoints-out:eq}
\end{eqnarray}%
\label{turnpoints-in_out:eq}%
\end{subequations}
where the upper/lower sign of $\mp$
distinguishes the first and second turning points.
Note that $\tau_{1}^{\rm out} < \tau_{2}^{\rm out}$, and the turning points
are real if either $m_{\rm eff}>0$ or $\varepsilon \ge - 2\, m_{\rm eff}$
for $m_{\rm eff} <0$.

For a given energy $E$ and momentum $p_\varphi$ one can calculate
the cyclotron radius $\varrho$ and the guiding center $c$:
\begin{subequations}
\begin{eqnarray}
\varrho & = & l\, \sqrt{\varepsilon},
\label{cycl_radius:eq} \\[1ex]
c & = & l \, \sqrt{\varepsilon + 2\, m_{\rm eff}}.
\label{guiding_center:eq}
\end{eqnarray}%
\label{rho_c:def}%
\end{subequations}
The derivation in the frame of classical mechanics is outlined
in  Appendix~\ref{guiding:app}.
Following  Ref.~\onlinecite{Lent:cikk} the relation between
these quantities and the corresponding quantum states of the system
can be derived from quantum mechanics.
It turns out that the same
relations hold for the cyclotron radius and the guiding center provided
$p_\varphi$ is quantized as $p_\varphi =\hbar m$, where $m$ now is an
integer.
Then, $m_{\rm eff}$ is the same as that defined by Eq.~(\ref{m_eff:def}).
The same results were found by Sim et al.~\cite{Sim:cikk}.
Using Eq.~(\ref{rho_c:def}) we shall discuss in detail
the correspondence between classical orbits and quantum states
in Secs.~\ref{phase_diag:sec} and \ref{current:sec}.

The turning points given in Eq.~(\ref{turnpoints-out:eq}) can be
expressed in terms of the cyclotron radius and the guiding center:
\begin{subequations}
\begin{eqnarray}
l\, \tau_{1}^{\rm out}   &=& \left\{ \begin{array}{ll}
c-\varrho, & \text{if} \,\,\, m_{\rm eff}>0,
\label{1st_turnig_guiding:eq} \\[1ex]
\varrho -c  , & \text{if} \,\,\, m_{\rm eff}< 0,
\end{array}
\right.         \\[1ex]
l\, \tau_{2}^{\rm out}   &=& c+\varrho.
\label{turnpoint_guiding:eq}
\end{eqnarray}
\end{subequations}

We can now classify the classical orbits according to the relation
between the values of the turning points given by 
(\ref{turnpoints-in_out:eq}) and the corresponding radius of the
circular non-magnetic region (in units of $l$)
\begin{equation}
\tau_{\rm R} = \frac{R}{l} = \sqrt{2s}.
\end{equation}
There are two different cases listed in Table \ref{class_cond:table}.
For orbits of type $A$ the particle outside the non-magnetic region
moves along a cyclotron orbit and then passes through the magnetic
field free region as a free particle.
In the case of orbits of type $B$ the particle does not penetrate the
non-magnetic region.
In this case one can further distinguish two additional types
of cyclotron orbits depending on the sign of $m_{\rm eff}$.
The condition $\tau_1^{\rm out} > \tau_{\rm R}$ listed
in Table~\ref{class_cond:table} and Eq.~(\ref{1st_turnig_guiding:eq})
imply that $c-\varrho > R$ for $m_{\rm eff} >0 $, and
$\varrho -c  > R$ for $m_{\rm eff} <0 $.
From a simple geometrical consideration it follows that
in the first case the cyclotron orbits (denoted by $B_1$) lie outside the
circle of radius $R$, while in the latter case the orbits (denoted by $B_2$)
completely encircle the non-magnetic region.
These conditions can be rewritten as
\begin{subequations}
\begin{eqnarray}
m\, {\rm sgn}(eB) & > & \sqrt{2\,s\,\varepsilon} > 0,
\,\,\, \text{for}\,\,\, B_1, \\
-m\, {\rm sgn}(eB) & > & \sqrt{2\,s\,\varepsilon} > 0,
\,\,\, \text{for}\,\,\, B_2.
\end{eqnarray}
\end{subequations}
For both types $B_1$ and $B_2$, $\varepsilon <  m^2/2s$ is valid.
In the case of orbits of type $A$ we have $\varepsilon > m^2/2s$.

We now turn to the calculation of the radial action
variable $I_r$. Using the radial potential
given by (\ref{Vm-pot:eq}) inside the non-magnetic circle we find
\begin{subequations}
\begin{eqnarray}
\lefteqn{\Theta_{\rm in} (\varepsilon,\tau )  \equiv
\frac{1}{\hbar}\, \int p_r \, d r
= \int \sqrt{\varepsilon -V(\tau) }  \, d \tau
}
\nonumber \\
& = &  \sqrt{\varepsilon\tau^2-m^2}
- m \, \arccos\, \frac{m}{\tau\,\sqrt{\varepsilon}},
\label{Theta_inside:eq}
\end{eqnarray}
and similarly outside,  we have
\begin{eqnarray}
\lefteqn{\Theta_{\rm out} (\varepsilon,\tau )  \equiv
\frac{1}{\hbar}\, \int p_r \, d r
= \int \sqrt{\varepsilon -V (\tau) }  \, d \tau
}
\nonumber \\
& = & \frac{1}{2}\,
\sqrt{\varepsilon\, \tau^2-\left(\frac{\tau^2}{2} - m_{\rm eff}\right)^2}
\nonumber \\
&& -\frac{1}{2}\, (\varepsilon + m_{\rm eff})
\arcsin\left(
\frac{\varepsilon + m_{\rm eff}-\tau^2/2}
{\sqrt{\varepsilon \left(\varepsilon + 2m_{\rm eff}\right)}}
\right)
\nonumber \\[2ex]
&& - \frac{1}{2}\, |m_{\rm eff}|\arcsin\left(
\frac{\tau^2(\varepsilon + m_{\rm eff})
-2 m_{\rm eff}^2}{\tau^2
\sqrt{\varepsilon \left(\varepsilon + 2m_{\rm eff}\right)}}
\right).
\label{Theta_outside:eq}
\end{eqnarray}
\end{subequations}
The radial action variables for the orbits of types $A$ and $B$ can be
expressed in terms of the functions
$\Theta_{\rm in}$ and $\Theta_{\rm out}$, and are listed
in Table \ref{class_cond:table}.
\begin{table}[hbt]
\caption{\label{class_cond:table}
Classification of the orbits and the corresponding radial action variables.
See also the text.  }
\begin{ruledtabular}
\begin{tabular}{lcr}
Case & conditions & $\frac{\pi}{\hbar}\, I_r$ \\[1ex] \hline
$A$ & $\tau_0^{\rm in} \le \tau_{\rm R}$ and &
$\Theta_{\rm out} (\varepsilon,\tau_2^{\rm out})
- \Theta_{\rm out} (\varepsilon,\tau_{\rm R})$         \\
&  $\tau_2^{\rm out} > \tau_{\rm R}$ &
$+\, \Theta_{\rm in} (\varepsilon,\tau_{\rm R}) -
\Theta_{\rm in} (\varepsilon,\tau_0^{\rm in})$   \\[2ex] \hline
$B$ & $\tau_1^{\rm out} > \tau_{\rm R}$ &
$\Theta_{\rm out} (\varepsilon,\tau_2^{\rm out})
-\Theta_{\rm out} (\varepsilon,\tau_1^{\rm out})$  \\[1ex]
\end{tabular}
\end{ruledtabular}
\end{table}

Note that
\begin{subequations}
\begin{eqnarray}
\Theta_{\rm out} (\varepsilon,\tau_1^{\rm out})
& = & -\frac{\pi}{4}\, \left( \varepsilon +m_{\rm eff}
- \left| m_{\rm eff}\right| \right),  \\
\Theta_{\rm out} (\varepsilon,\tau_2^{\rm out})
& = & - \Theta_{\rm out} (\varepsilon,\tau_1^{\rm out}).
\end{eqnarray}
\end{subequations}
Thus, for orbits of type $B$,
the radial action variable $I_r$ can be simplified to
\begin{equation}
I_r = \frac{\hbar}{2}\, \left( \varepsilon +m_{\rm eff}
- \left| m_{\rm eff}\right| \right).
\label{I_r_B-type:eq}
\end{equation}

It is clear from (\ref{angular_action:eq}) and Table~\ref{class_cond:table}
that for fixed $s$, the radial action variable
$I_r$ is a function of the rescaled energy $\varepsilon$ and
the angular action variable $I_\varphi$ through $m$ and $m_{\rm eff}$.
Then, for orbits of type $A$, the partial derivative in the
denominator of Eq.~(\ref{BerryTabor2d}) has a rather simple form
\begin{equation}
\hbar \, \frac{\partial^2 I_r }{\partial I_\varphi^2}
= \frac{1}{\pi\, \sqrt{2\,s\,\varepsilon - m^2}}\,
\frac{s + m_{\rm eff}}{\varepsilon +2\,m_{\rm eff} }.
\label{eq:g2I2}
\end{equation}

However, the amplitude for orbits of type $B$
in Eq.~(\ref{BerryTabor2d}) cannot
be calculated using the second partial derivative of $I_r$,
therefore the contribution from these orbits to the semiclassical level
density is calculated separately in Appendix~\ref{Landau:app}.

Knowing the explicit $\varepsilon$ and $I_\varphi$ dependence
of the radial action variable $I_r$,
the Bohr-Sommerfeld quantization conditions given by Eq.~(\ref{action})
for orbits of type $A$ can be rewritten as
\begin{subequations}
\begin{eqnarray}
I_\varphi &=& \hbar \, m, \\
I_r &=& \hbar \left(n+\frac{1}{2}\right),
\end{eqnarray}
where $n=0,1,2,\cdots$ and $m$ is an integer, and the energy-dependent
radial action variable $I_r$ is given in Table~\ref{class_cond:table}
(the Maslov indices are $\nu_\varphi =0$ and $\nu_r = 2$,
for details see, eg, Ref.~\onlinecite{Brack:book}).
Using (\ref{I_r_B-type:eq}) for orbits of type $B$,
the semiclassical quantization conditions can be simplified
and the energy levels are
\begin{equation}
E_{m,n} = \hbar\, \omega_c \,
\left(n+ \frac{\left|m_{\rm eff} \right|-m_{\rm eff} +1}{2}\right),
\end{equation}%
\label{BS_quant:eq}%
\end{subequations}
where $m_{\rm eff}= s+m\, {\rm sgn}(eB)$, and $m$ and $n$ are integers.
These levels coincide with the familiar Landau levels
in a homogeneous magnetic field
but the quantum number $m$ is replaced by $m_{\rm eff}$.
Below in Sec.~\ref{Berry_Tabor_results:sec} we shall compare
the exact energy levels with those obtained from the Bohr-Sommerfeld
quantization conditions for orbits of types $A$ and $B$.

\subsection{Periodic orbits}\label{periodic_orbit:sec}

\begin{figure}[t!]
\centerline{\includegraphics[width=0.85\columnwidth]{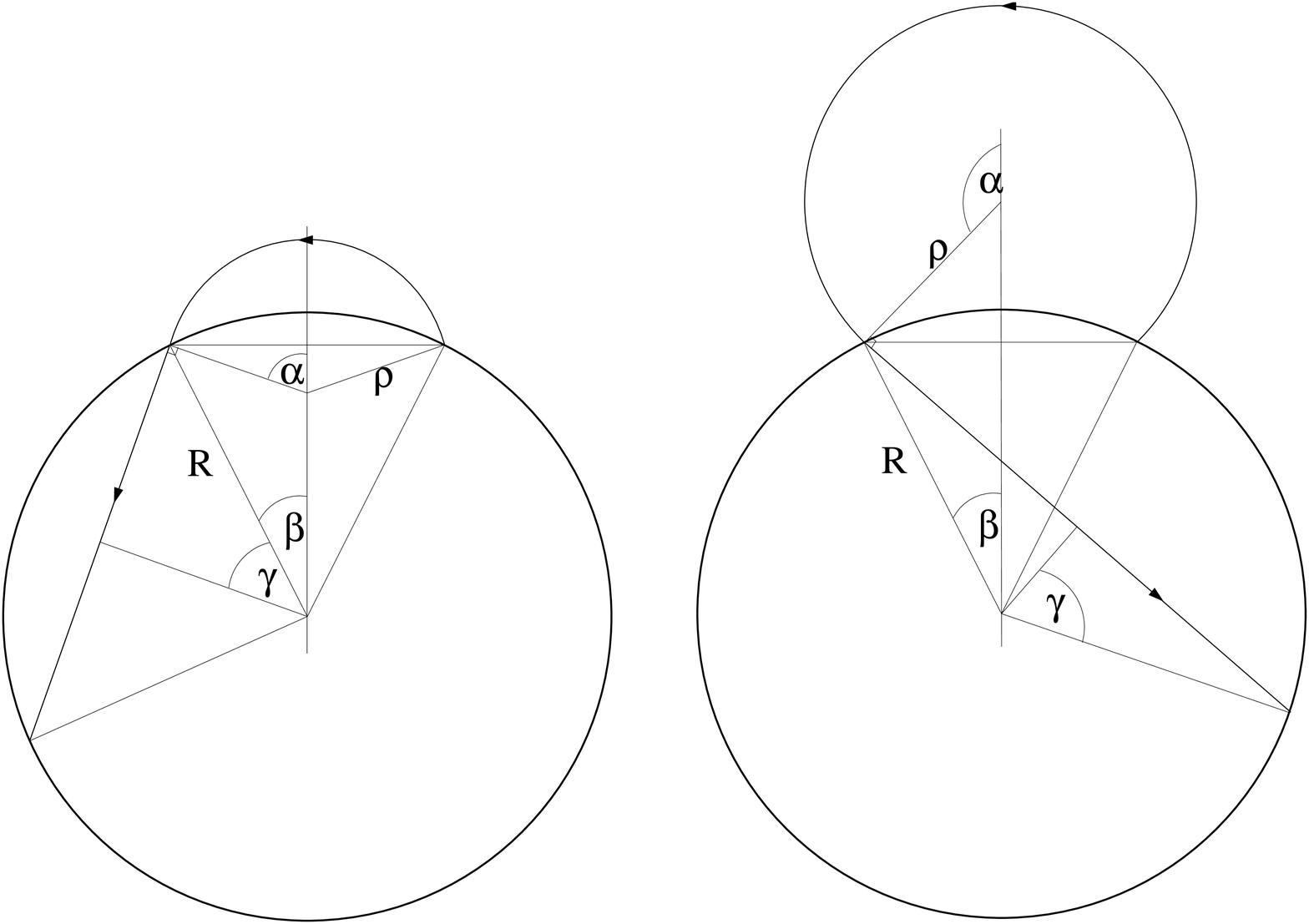}}
\caption{Two examples for the basic orbit segments (an arc in the
magnetic field followed by a straight line inside the anti-dot) of
orbit type $A$, and  the related angles $\alpha,\beta$ and $\gamma$.
A particle traveling along the segments moves anti-clockwise with
respect to the center of the anti-dot in the case drawn on the
left, while in the other example it moves clockwise on the straight
line inside the anti-dot. Thus in this case the angle $\gamma$ is
negative. }
\label{fig:szogek}
\end{figure}

To apply the Berry-Tabor formula (\ref{BerryTabor2d}), one needs to
describe the possible periodic orbits of the magnetic anti-dot
system.
The periodic orbits of type $A$ can be characterized by their winding
number $w$ (the number of turns around the center under one cycle) 
and the number of identical orbit segments $n_s$ the orbit can be
split up to. These segments consist of a circular path outside 
the anti-dot followed by a straight line inside. 
We introduce the angles $\alpha,\beta$ and $\gamma$ to characterize 
these basic orbit segments as shown in Fig. \ref{fig:szogek}.
These angles always fulfill
\begin{subequations}
\begin{eqnarray}
n_s(\beta+\gamma)&=&w\pi,
\label{cond_po:eq}       \\
\frac{\sin(\alpha)}{\sin(\beta)}&=&\frac{R}{\rho},
\end{eqnarray}
\end{subequations}
($\alpha$ and $\beta$ are always positive and the sign of $\gamma$
follows the sign of $w$).
The relations between the indices $w,n_s$ and the angles characterizing
the basic orbit segment are summarized in Table \ref{table:orbits}
for the four possible sub-classes.
When either $w$ is negative (orbits of type $A{}_1$), or the cyclotron
radius $\rho$ is smaller than the radius of the anti-dot
(orbits of type $A{}_2$), the
angles $\alpha,\beta$ and $\gamma$
are  fully determined by $w$ and $n_s$, since in these cases
$\beta$ is definitely smaller than $\pi/2$.  On the other hand,
when $w>0$ and $\rho>R$, one must also specify whether $\beta$ is
smaller (orbits of type $A{}_3$) or larger (orbits of type $A{}_4$)
than $\pi/2$ to fully determine the periodic orbit.

\begin{table}[hbt]
\caption{\label{table:orbits}
The different sub-classes of orbits of type $A$ and the corresponding
relations between the angles defining the basic orbit segment
and the indices $w$ and $n_s$. Every orbit with a negative winding
number $w$ falls into sub-class $A{}_1$, regardless of whether $\rho$ is
smaller or larger than $R$. The sub-class $A{}_2$ consists of
those orbits which have a positive winding number and $\rho<R$.
In case of $w>0$, $\rho>R$ and $\beta<\pi/2$, the orbit is of
type $A{}_3$, otherwise it is of type $A_{4}$. The angle $\alpha$
can be obtained from $w$ and $n_s$ directly. Then $\beta$ can
be calculated from $\alpha,\rho$ and $R$, and finally $\gamma$ from
$\alpha$ and $\beta$.}
\begin{tabular}{|c||c|c|c|}
\hline  $A_1:\;\; w<0$
& $A_2: \left\lbrace\begin{matrix} w>0, \cr
\rho< R\end{matrix}\right.$
&$A_3:\left\lbrace \begin{matrix} w>0, \cr \rho>R,
\cr \beta<\pi/2\end{matrix}\right.$
&  $A_4: \left\lbrace\begin{matrix} w>0, \cr \rho>R,\cr
\beta>\pi/2\end{matrix}
\right.$ \\
\hline
{\raisebox{-0.5cm}{\includegraphics[scale=0.15]{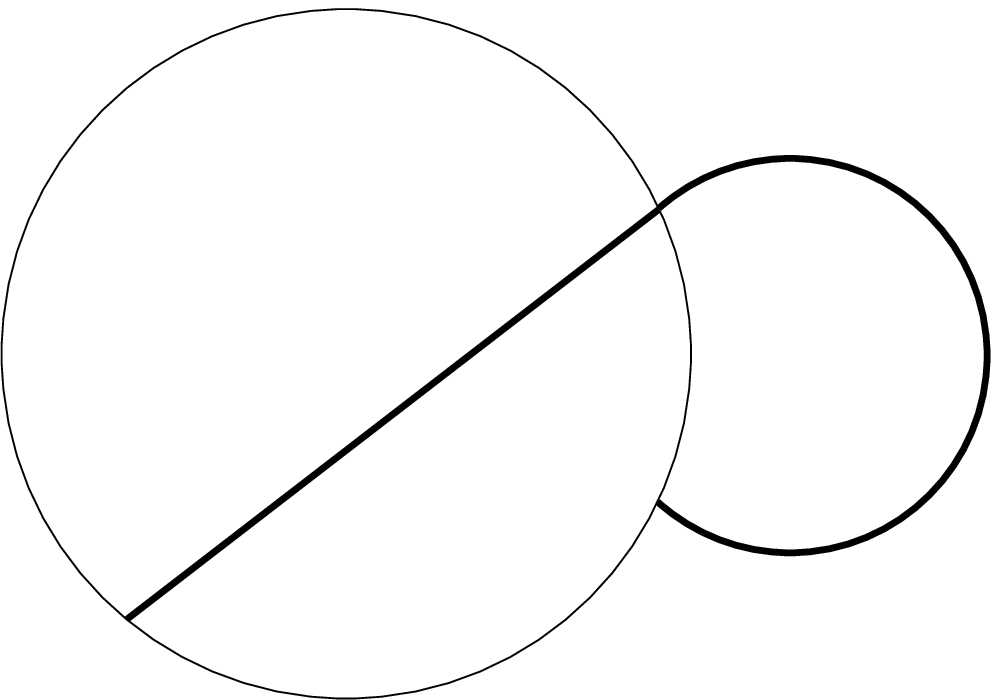}}}
& $\;\;${\raisebox{-0.5cm}{\includegraphics[scale=0.15]{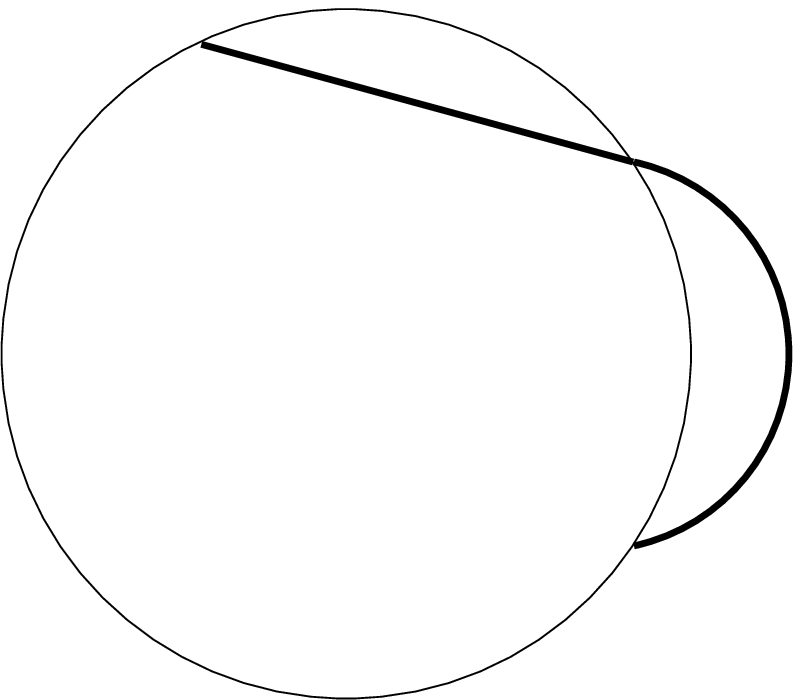}}}
&{\raisebox{-0.55cm}{\includegraphics[scale=0.15]{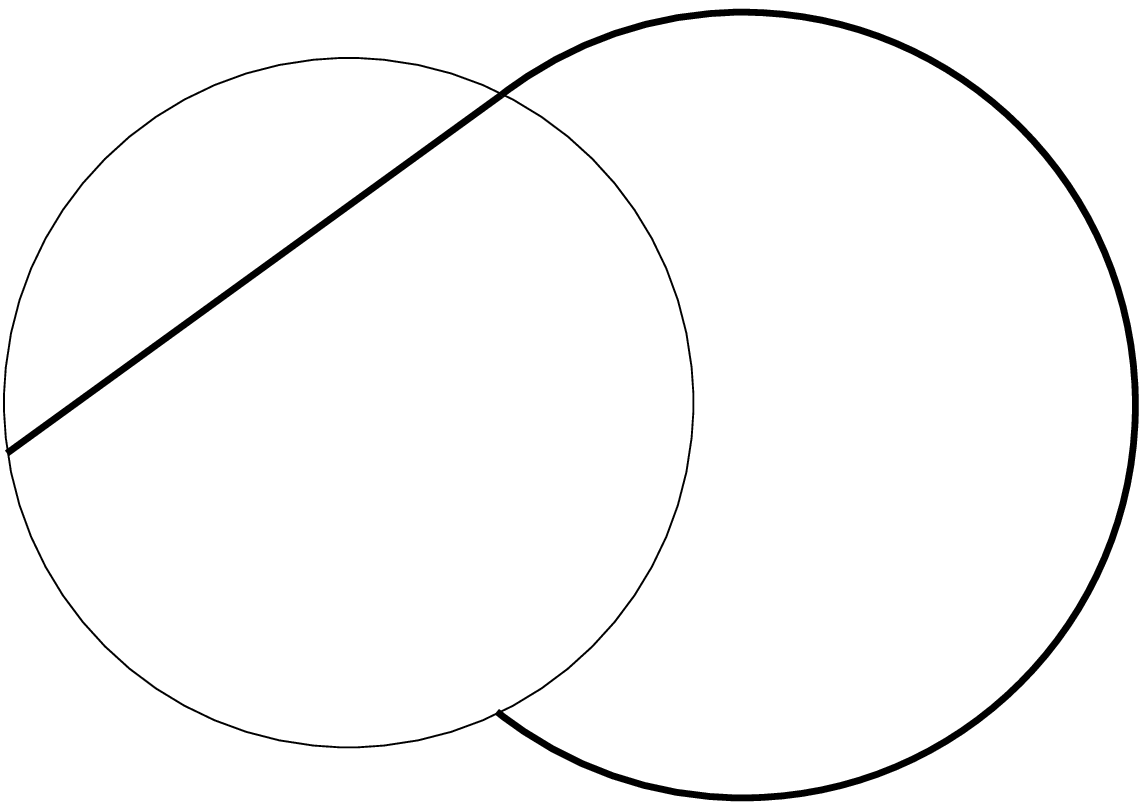}}}
& $\begin{matrix} \cr  \cr \cr 
\end{matrix}\;\;${\raisebox{-0.55cm}{\includegraphics[scale=0.15]{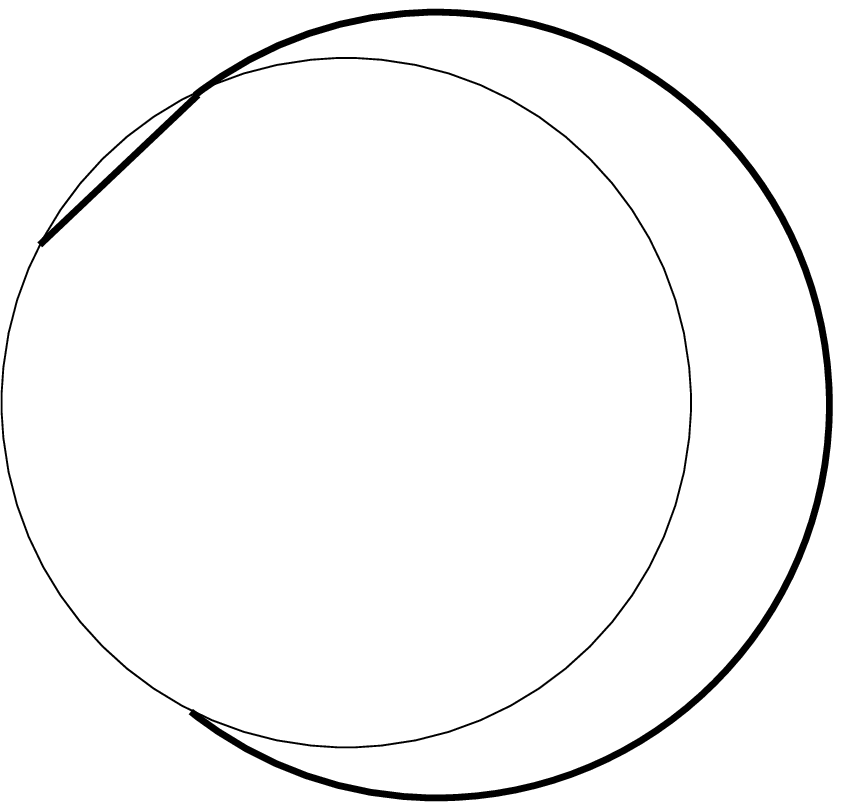}}} 
  \\
\hline
$\alpha=\pi+\frac{\pi w}{n_s}$ & \multicolumn{3}{c|}{
$\alpha=\frac{\pi w}{n_s}$} \\
\hline
\multicolumn{3}{|c|}{$\beta=\beta_0\equiv\arcsin(\frac{\rho}{R}\sin 
\alpha)$}
&$\beta=\pi-\beta_0$\\
\hline
$\gamma=\alpha-\pi-\beta$ & \multicolumn{3}{c|}{
$\gamma=\alpha-\beta$}\\
\hline
\end{tabular}
\end{table}

The action of periodic orbit $p$ can generally be expressed as
\begin{equation}
S_p= \hbar kL_p + eB{\cal A}_p,
\label{action_general:def}
\end{equation}
where $k=\sqrt{2ME}/\hbar$ is the wave number,
$L_p$ is the length of the orbit and
${\cal A}_p$ is the area inside the magnetic field. In our case
\begin{eqnarray}
L_p&=&2n_s\left[\rho\, \alpha +R\sin(|\gamma|)\right],\\
{\cal A}_p&=&n_s\left\{\rho^2\left[\alpha +
{\rm sgn}\left(2\alpha-\pi\right)\,
\frac{ \sin 2\alpha }{2}\right]
\right. \nonumber \\ & &\left.
-R^2\left[\beta +
{\rm sgn}\left(2\beta-\pi\right)\,
\frac{ \sin 2\beta }{2}\right]
\right\} .
\end{eqnarray}
Therefore the action in our units can be written as
\begin{subequations}
\begin{eqnarray}
S_p/\hbar&=&
n_s\left\lbrace 2\, \varepsilon \, \alpha +
2\sqrt{2s\,\varepsilon}\, \sin(|\gamma|)
\right.\nonumber \\ & &
+{\rm sgn}(eB)  \left[ \varepsilon \left(\alpha +
{\rm sgn}\left(2\alpha-\pi\right)\,
\frac{ \sin 2\alpha }{2}\right) \right.
\nonumber \\ & &
\left.  \left.
-2s \left(\beta +
{\rm sgn}\left(2\beta-\pi\right)\,
\frac{ \sin 2\beta }{2}\right)\right]
\right\rbrace.
\label{action_A}
\end{eqnarray}
Finally, to use (\ref{BerryTabor2d}) and
(\ref{eq:g2I2}),
we also need the time period $T_p$ and $I_{\varphi}$ associated to the 
orbit,
which can be written as
\begin{eqnarray}
T_p&=&\frac{L_p}{v}=\frac{2n_s}{\omega_c}\left(1+\sqrt{2s/\varepsilon}
\sin(|\gamma|)\right),\\
I_{\varphi,p}/\hbar&=&\pm 
\sqrt{2s}\varepsilon\cos(\gamma),\label{I_phi_orbit}
\end{eqnarray}%
\label{eq:periodic_orbit_props}%
\end{subequations}
where $v$ denotes the velocity and in the latter expression the upper sign
is for the orbits with $w>0$ and the lower sign is
for the orbits with $w<0$. This expression for $I_{\varphi}$ can be obtained
from (\ref{angular_action:eq}): the angular action variable is equal
to $p_{\varphi}$, and, as already mentioned, inside the anti-dot
$p_{\varphi}$ is equivalent to the angular momentum.

\subsection{Results}\label{Berry_Tabor_results:sec}

In this section we compare the numerically exact energy levels with
those calculated from the Bohr-Sommerfeld quantization conditions.
Similarly, we present  results for the density of states obtained from the
Berry-Tabor formula (\ref{BerryTabor2d}).

\begin{figure}[hbt]
\includegraphics[scale=0.7]{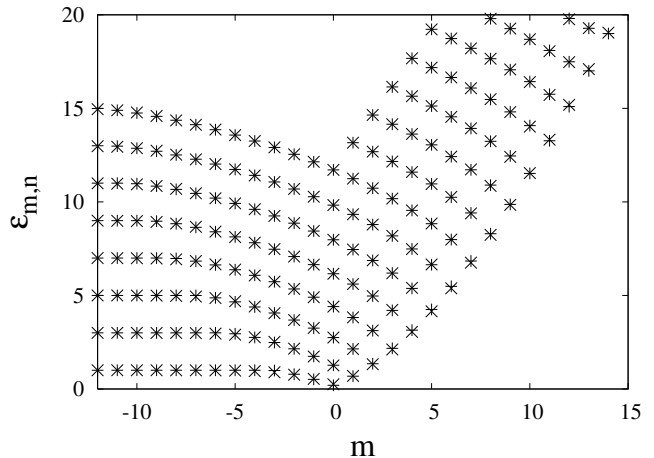}
\caption{Exact (crosses) and semiclassical (+ signs) energy levels
(in units of $\hbar \omega_c/2$) of the
circular magnetic anti-dot obtained from Eqs.~(\ref{secular:eq}) and
(\ref{BS_quant:eq})  as functions of $m$ for $s=5$.
We take $\rm{sgn}(eB) = -1$ as in Ref.~\onlinecite{Sim:cikk}.
\label{exact_levels-BS:fig}}
\end{figure}

The numerically exact energy levels of the magnetic anti-dot system are
calculated from the secular equation (\ref{secular:eq}) for fixed $m$.
Solving Eq.~(\ref{BS_quant:eq}) for $\varepsilon$
we obtain the energy levels in the Bohr-Sommerfeld approximation.
The results for a given magnetic field are shown
in Fig.~\ref{exact_levels-BS:fig}.
The agreement between the exact and the semiclassically
calculated energy levels is excellent.
Our results also agree with those presented in Ref.~\onlinecite{Sim:cikk}.
For large $|m|$ the energy levels tend to the Landau levels,
while in the opposite case a substantial deviation can be
seen.
In the latter case, the energy levels result from the quantization
of orbits of type $A$. 
In the work by Sim et al.~\cite{Sim:cikk} these states were called magnetic
edge states.
One can see that even the low-lying energy levels of these 
magnetic states can be accurately calculated 
in the Bohr-Sommerfeld approximation.
However, a significant deviation of the eigenvalues of these states
from the bulk Landau levels can be seen in the figure.
The lowest energy level of the magnetic anti-dot system is the state
$m=0$ and $n=0$.
Note that the spectrum can be calculated much more easily in the
semiclassical approximation than from the exact secular equation involving
the confluent hypergeometric function $U$.

Increasing the magnetic field, we experienced slight deviations.
These discrepancies may be explained qualitatively in the following way.
As the magnetic field tends to infinity, the charged particle
spends less and less time outside the circular region,
and in the limiting case its motion is
described by an elastic reflection from the boundary of the magnetic and
non-magnetic regions.
The radial potential becomes a hard wall at $r=R$.
Thus, one of the classical turning points
for orbits of type $A$ becomes a
hard one and the corresponding contribution to the
Maslov index tends to 2.
Blaschke and Brack~\cite{Blaschke:cikk} observed a similar situation
in circular magnetic billiards. 
Their numerical investigations have confirmed the argument presented above.
Here we do not discuss this issue further.

\begin{figure}[hbt]
\includegraphics[scale=0.43]{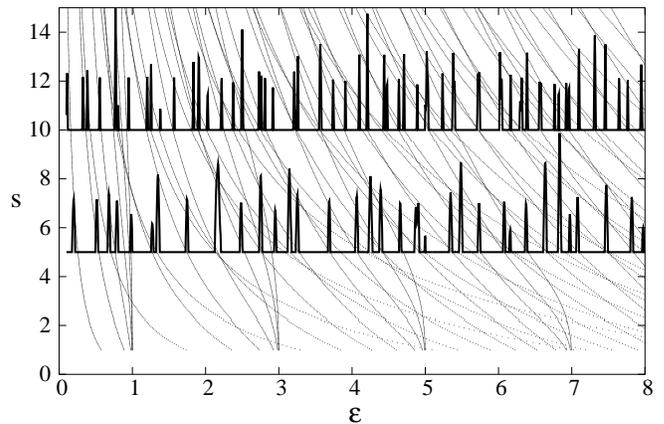}
\caption{The quantum mechanical energy levels as a function of $s$,
and the semiclassical level density obtained from the Berry-Tabor
approach as function of $\varepsilon$ are plotted together.
The smoothing parameter $\delta$
for the periodic orbit sum was $\delta=0.002$, and the summation of the
orbits runs from $n_s=2$ to $n_{s,\rm max}=240$ and from $w=0$ to
$w_{\rm max}=120$. 
(Numerical experience showed that with these
maximum values $n_{s,\rm max}$ and $w_{\rm max}$ the broadened
delta functions are quite prominent at semiclassical energies ).
\label{fig:Berry-QM}}
\end{figure}

We now present results for the density of states calculated from the
Berry-Tabor formula (\ref{BerryTabor2d}) for two magnetic fields
given by the missing flux quanta $s=5$ and $s=10$.
To evaluate the semiclassical density of states in
practice, we have regularized the periodic orbit sum in (\ref{BerryTabor2d})
with a Gaussian smoothing by multiplying the amplitude of
the orbits with  $e^{-\delta L_p}$, (where $\delta$ is infinitesimal),
as discussed in~\cite{Blaschke:cikk,Brack:book}.
This factor suppresses the contribution from
the long orbits and broadens the delta functions at semiclassical energies.
Substituting Eqs.~(\ref{eq:g2I2}) and (\ref{eq:periodic_orbit_props})
into the regularized version of (\ref{BerryTabor2d}),
we obtain the semiclassical density shown
in Fig.~\ref{fig:Berry-QM} plotted together with the
numerically obtained quantum energy levels.
The agreement between the two results is good for the majority of
the levels, however, in the case of missing flux quanta $s=10$,
apparent discrepancies can be observed for eaxample at energies
close to $\varepsilon \approx 3$.
We think that a better agreement can be obtained for stronger magnetic
fields by taking into account the magnetic field dependent
Maslov index. The work along this line is in progress.

\subsection{Phase diagram, the classical-quantum correspondence}
\label{phase_diag:sec}

In this section we classify the exact energy levels in terms of
the classical orbits.
In  Sec.~\ref{class_dyn:sec} orbits of  types $A$ and $B$ have been
introduced according to the positions of the turning points compared to
the magnetic anti-dot.
Orbits of type $A$ can further be classified into sub-classes $A_1$--$A_4$,
as it has been shown in Sec.~\ref{periodic_orbit:sec}.
However, in that section only periodic orbits have been studied.
We now go beyond the condition (\ref{cond_po:eq}) for periodic orbits. 
Then the four types of sub-classes can be characterized
by the angles $\alpha, \beta$ and $\gamma$, the cyclotron radius
$\varrho$ and the guiding center $c$ of the classical orbits.
Using simple geometrical arguments, these angles can be calculated
from $\varrho$ and $c$.
Using Eq.~(\ref{rho_c:def}) (which is valid in the quantum case, too)
the above classical parameters classifying the orbits can be directly
related to the quantum states given by the quantum number $m,n$ (and so
the energy eigenvalue $E_{m,n}$) and the missing flux quanta $s$.
Thus, the conditions given in the first row of Table~\ref{table:orbits}
for the different types of orbits can be reformulated
in terms of the particle energy $\varepsilon$
and the quantum number $m$.
The results are summarized in Table \ref{table:orbits_A-B-conds}.
\begin{table}[hbt]
\caption{\label{table:orbits_A-B-conds}
Conditions in terms of $\varepsilon$ and $m$ for
different sub-classes of orbits of types $A$ and $B$.
Here we take ${\rm sgn}(eB) = -1$. }
\begin{tabular}{|c|c|c|c|c|c|}
\hline
${\rm A}_1$ & ${\rm A}_2$ & ${\rm A}_3$ & ${\rm A}_4$ & $B_1$ & $B_2$
\\[1ex] \hline
\multicolumn{4}{|c|} {$\varepsilon > m^2/2s$ } &
\multicolumn{2}{|c|} {$\varepsilon \le m^2/2s$ }   \\[1ex] \hline
\raisebox{-0.55cm}{$m< 0$} &
\multicolumn{3}{|c|} {\raisebox{-0.15cm}{$ m \ge 0 $ }}
& \raisebox{-0.55cm}{$m \le 0$} & \raisebox{-0.55cm}{$m> s$}
\\ \cline{2-4}
&  \raisebox{-0.2cm}{$\varepsilon < 2s$} &
\multicolumn{2}{|c|} {$\varepsilon \ge 2s  $ } &&
\\[1ex] \cline{3-4} \cline{6-6}
& & $ m \le  2s $ & $ m > 2s $ && $\varepsilon \ge 2(m-s) $
\\[1ex] \hline
\end{tabular}
\end{table}
A similar classification has been made for electronic
states of a circular ring in magnetic field~\cite{Klama}
and for ring-shaped Andreev billiards~\cite{Bdisk:cikk}.

\begin{figure}[hbt]
\includegraphics[scale=0.7]{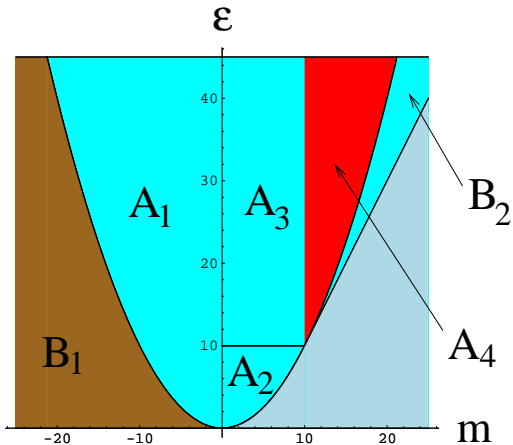}
\caption{Phase diagram in the space of $\varepsilon$ and $m$:
the classically allowed regions defined by
the conditions in Table~\ref{table:orbits_A-B-conds}
for the different orbits. Here $s=5$ and ${\rm sgn}(eB) = -1$.
\label{fazis_diag:fig}}
\end{figure}

It may also be useful to present the conditions listed 
in Table \ref{table:orbits_A-B-conds} graphically.
In $\varepsilon$-$m$ space, the classically allowed regions
corresponding to the different orbits look like a `phase diagram'.
In Fig.~\ref{fazis_diag:fig} such a phase diagram is plotted
in the space of $\varepsilon$ and $m$
In $\varepsilon$-$m$ space,
for a given magnetic field.
This phase diagram should be compared
with Fig.~\ref{exact_levels-BS:fig}, the plot of the energy levels 
from the quantum mechanical calculation.
In this way, the different exact levels can be classified in terms of the
corresponding classical orbits.
We should stress again that these orbits are not necessarily periodic.
Examples will be shown in Sec.~\ref{current:sec}.

\section{Comparison of the current distributions and the classical
trajectories}
\label{current:sec}

An apparent correspondence between the classical orbits and the quantum 
states can also be made by calculating the current flow
patterns in the magnetic anti-dot system.
The particle (probability) current density~\cite{Lent:cikk,Schwabl:book}
in magnetic field is given by
\begin{equation}
{\bf j} = \frac{i\hbar}{2M} \left(\Psi \, {\rm grad} \Psi^* - \Psi^* {\rm
grad} \Psi\right) - \frac{e}{M}\, {\bf A} {|\Psi|}^2.
\label{current:def}
\end{equation}
Using the vector potential (\ref{vector-pt:eq})
the current density (in our units) for states $\Psi_{m,n}$ can be written
as
${\bf j} = j_\varphi \, \frac{\hat{{\bf e}}_\varphi}{r}$, where
\begin{equation}
j_\varphi(\tau) = \frac{\hbar}{2M}\, {|\Psi_{m,n}|}^2 \left[2m
-{\rm sgn}(eB)\left(\tau^2 -\tau_{\rm R}^2 \right)
\Theta\left(\tau -\tau_{\rm R}\right)\right],
\label{current:eq}
\end{equation}
and $\hat{{\bf e}}_\varphi$ is the unit vector in the $\varphi$ direction.

\begin{figure}[hbt]
\includegraphics[scale=0.3]{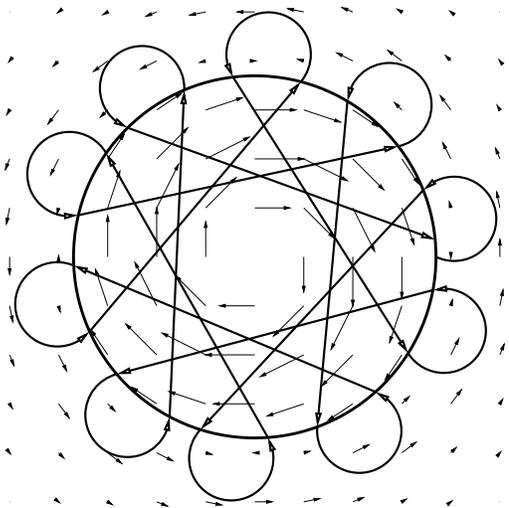}
\caption{The current flow (in units of $\hbar/(2M)$) 
for state $m=-1$ and $n=0$.
The classical orbit is of type $A_1$.
Here and in Figs.~\ref{aram_n0:fig}-\ref{aram_A3_A4:fig}
the missing flux quanta is $s=5$, and we choose ${\rm sgn}(eB) = -1$.
\label{nyilas_0_-1:fig}}
\end{figure}

\begin{figure}[hbt]
\includegraphics[scale=0.32]{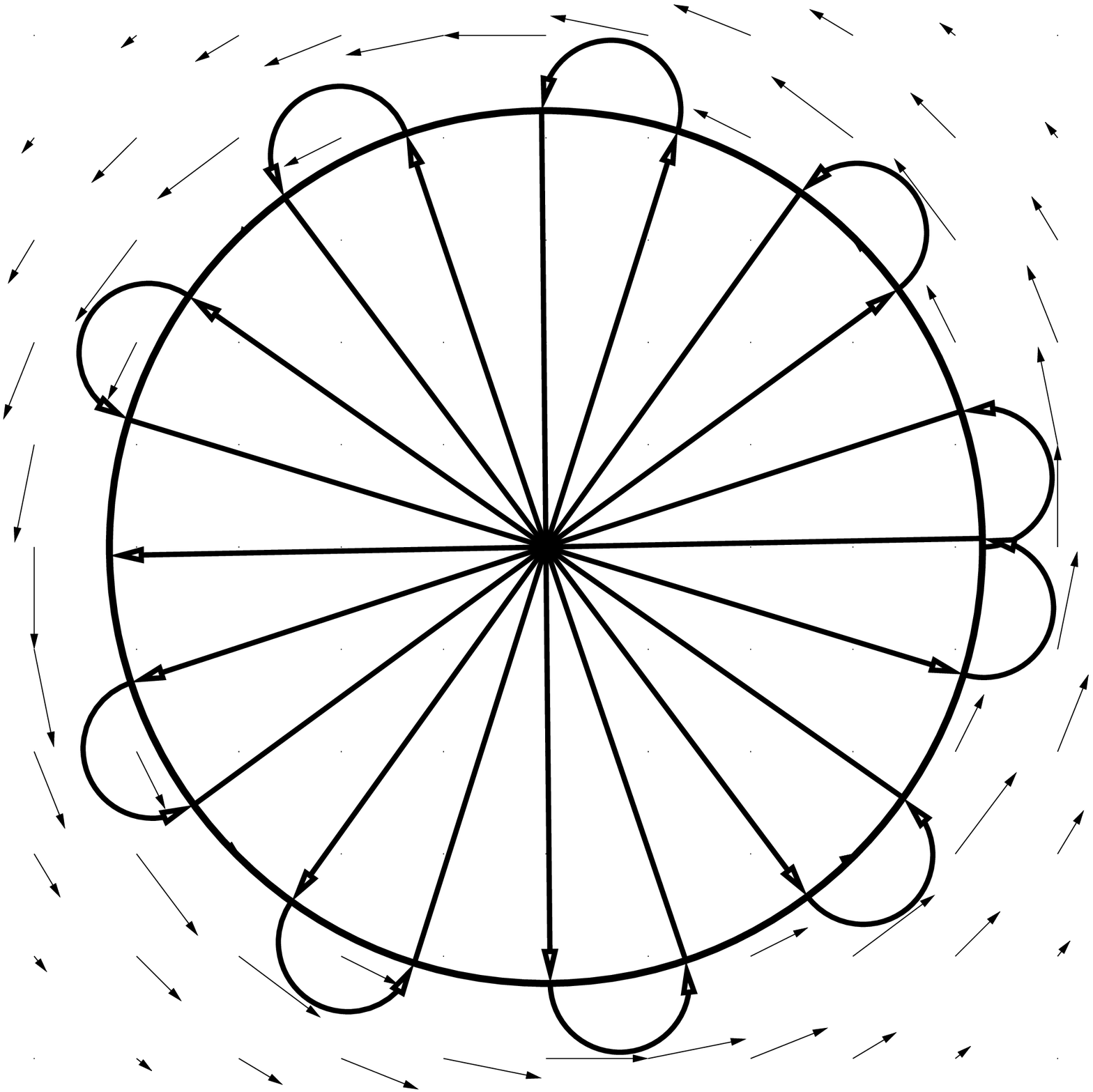}
\caption{The current flow (in units of $\hbar/(2M)$) 
for state $m=0$ and $n=0$.
The classical orbit is of type $A_2$.
\label{nyilas_0_0:fig}}
\end{figure}

\begin{figure}[hbt]
\includegraphics[scale=0.33]{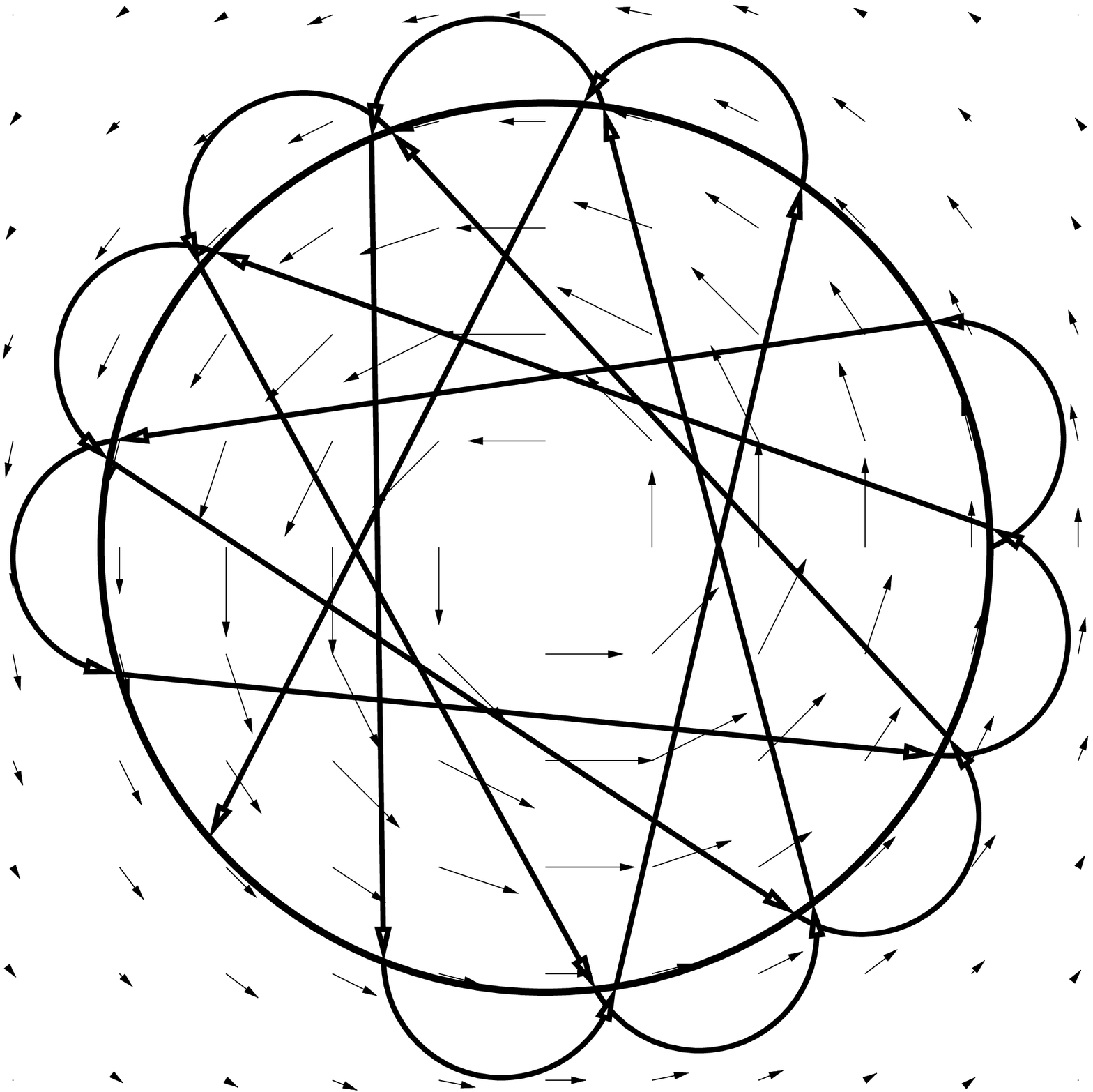}
\caption{The current flow (in units of $\hbar/(2M)$) 
for state $m=1$ and $n=0$.
The classical orbit is of type $A_2$.
\label{nyilas_0_1:fig}}
\end{figure}

\begin{figure}[hbt]
\includegraphics[scale=0.5]{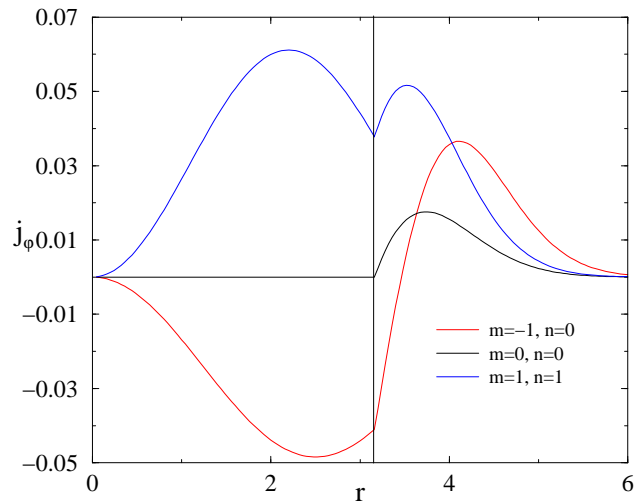}
\caption{The current densities $j_\varphi$ as functions of $r$ (in
units of $l$) for states shown 
in Figs.~\ref{nyilas_0_-1:fig}-\ref{nyilas_0_1:fig}.
The vertical line is at $r= R$.
\label{aram_n0:fig}}
\end{figure}

\begin{figure}[hbt]
\includegraphics[scale=0.17]{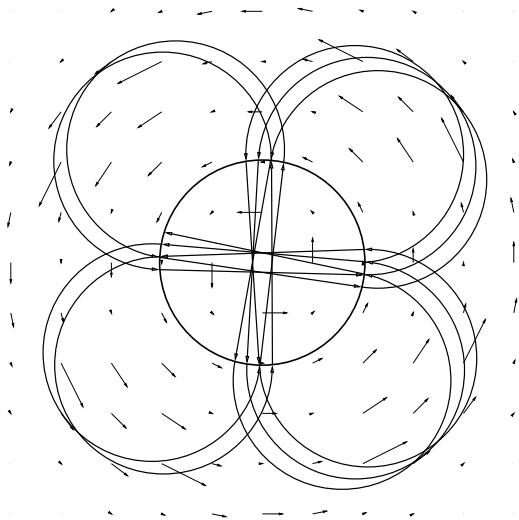}
\caption{The current flow (in units of $\hbar/(2M)$) 
for state $m=1$ and $n=6$.
The classical orbit is of type $A_3$.
\label{nyilas_A3:fig}}
\end{figure}

\begin{figure}[hbt]
\includegraphics[scale=0.17]{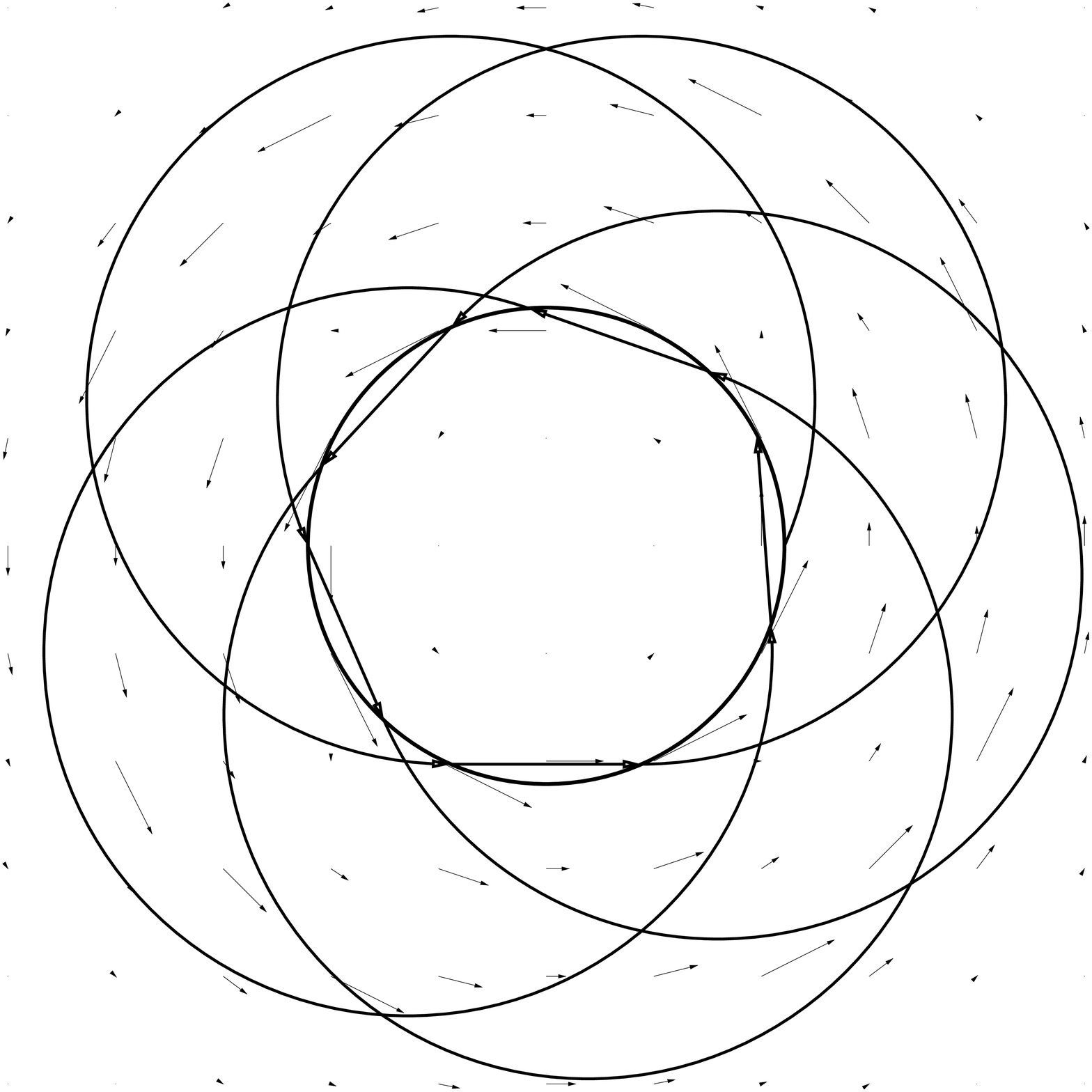}
\caption{The current flow (in units of $\hbar/(2M)$) for state
$m=14$ and $n=2$.
The classical orbit is of type $A_4$.
\label{nyilas_A4:fig}}
\end{figure}

\begin{figure}[hbt]
\includegraphics[scale=0.5]{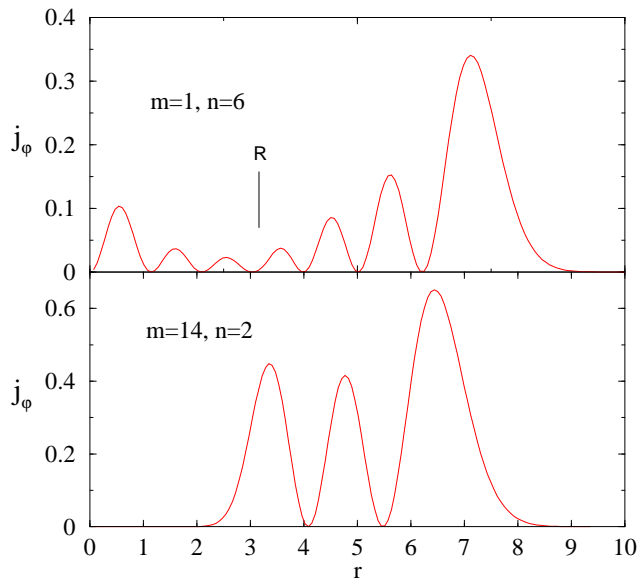}
\caption{The current densities $j_\varphi$ as functions of $r$ (in
units of $l$) for states shown
in Figs.~\ref{nyilas_A3:fig}-\ref{nyilas_A4:fig}.
The parameters are the same as in  Fig.~\ref{nyilas_0_-1:fig}.
The vertical line is at $r= R$.
\label{aram_A3_A4:fig}}
\end{figure}

Solving the secular equation (\ref{secular:eq}) and then determining the
normalized eigenstates, the related current densities can be calculated
from (\ref{current:eq}).
Figures~\ref{nyilas_0_-1:fig}-\ref{nyilas_0_1:fig},
\ref{nyilas_A3:fig} and~\ref{nyilas_A4:fig}
show the current flow patterns for given eigenstates and the corresponding
classical trajectories of the particle.
The missing flux quanta is $s=5$ and ${\rm sgn}(eB) = -1$ in all figures.
In these figures the current density ${\bf j}({\bf r})$ at point ${\bf r}$
is represented by an arrow with length proportional to
the magnitude of the current density and the midpoint of the arrow
is at the point ${\bf r}$.
At a given energy and quantum number $m$ corresponding
to the classical canonical momentum $p_\varphi = \hbar m$
the cyclotron radius and the guiding center of the classical orbit is
calculated from Eq.~(\ref{rho_c:def}).
Hence the classical trajectory of the particle can be determined
and are shown in the figures (scaling is in units of the magnetic length $l$).
In Figs.~\ref{aram_n0:fig} and~\ref{aram_A3_A4:fig}
the radial dependence of the current density is plotted 
for the corresponding eigenstates.
States $(m,n)= (-1,0), (0,0)$ and $(1,0)$ were called magnetic edge
states in Ref.~\onlinecite{Sim:cikk}.

In Fig.~\ref{nyilas_0_-1:fig} the current inside the non-magnetic region
flows clockwise (the magnitude of the current density is negative in
accordance with Fig.~\ref{aram_n0:fig}), while outside the magnetic
dot it flows counterclockwise.
The classical trajectories inside the dot form
a `caustic' and the current is enhanced here in the clockwise direction.
From Table~\ref{table:orbits_A-B-conds} we find that 
the orbit is of type $A_1$.
The trajectory apparently also satisfies the conditions given in
Table~\ref{table:orbits} (without the requirement for the periodicity of
the orbits).

In Fig.~\ref{nyilas_0_0:fig} for state $m=0$ and $n=0$ the current is
zero inside the magnetic dot.
This can easily be seen from Eq.~(\ref{current:eq}),
while the direction of the current flow is counterclockwise outside.
The orbit is a limiting case between types $A_1$ and $A_2$.
Figure~\ref{nyilas_0_1:fig} for state $m=1$ and $n=0$ shows 
a counterclockwise current flow both inside and outside the magnetic dot.
The current $j_\varphi$ is positive everywhere as it can be also seen
in Fig.~\ref{aram_n0:fig}.
The orbit is again of type $A_2$, in accordance with the conditions 
given in Tables~\ref{table:orbits} and~\ref{table:orbits_A-B-conds}.

One can observe that the current density $j_\varphi$ is not
differentiable at $r=R$. 
This is because the theta function in Eq.~(\ref{current:eq}). 
Physically, this is a consequence of the step function 
behaviour of the magnetic field. 
Nevertheless, the divergence of the current density vector ${\bf j}$ 
is still zero everywhere.

Finally, orbits of types $A_3$ and $A_4$ are shown
in Figs.~\ref{nyilas_A3:fig} and~\ref{nyilas_A4:fig} for states
$(m,n)= (1,6)$ and $(14,2)$, respectively.
Similarly, for these states the corresponding current densities
as functions of the distance from the origin are plotted
in Fig.~\ref{aram_A3_A4:fig}.
For both states the current is very small inside the magnetic dot.
In the case of state $(m,n)= (1,6)$, the trajectories almost
cross the origin ($m$ is small), while for state $(m,n)= (14,2)$
only a small portion of the trajectory penetrates into the magnetic 
anti-dot regions.
The qualitative agreement between the current flow patterns and
the classical trajectories is, again, clearly visible.

Note that in all of these figures the classical trajectories are not
periodic orbits with energy corresponding to the given
eigenstate.
This is not surprising, since the quantization should not be applied
to periodic orbits in real space but to the motion on the two-dimensional 
torus parametrized by the action variables and their canonically 
conjugate angle variables (for details see, eg, Ref.~\onlinecite{Brack:book}).
In fact, the Berry-Tabor formula (\ref{BerryTabor2d})
suggests that an infinite number of periodic orbits with proper weights
can only result in the correct quantum mechanical density of states.

\begin{figure}[hbt]
\includegraphics[scale=0.5]{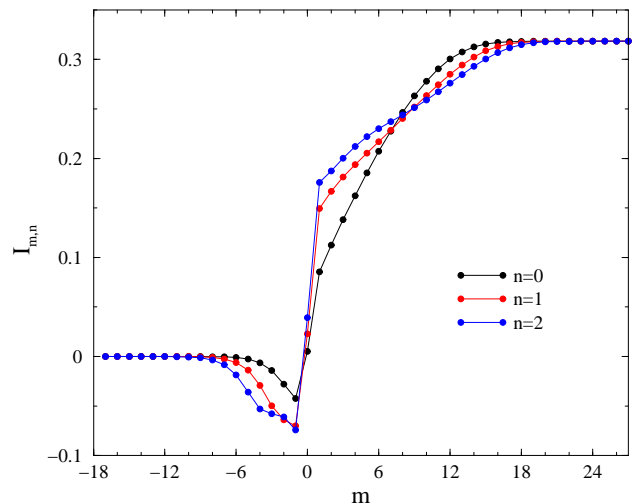}
\caption{The total current $I_{m,n}$ (in units of $\hbar/(2M)$)
obtained from the integral of $j_\varphi$ given in
(\ref{Halperin_current:def})
as functions of $m$ for $n=0, 1, 2$.
The parameters are the same as in  Fig.~\ref{nyilas_0_-1:fig}.
\label{total_aram:fig}}
\end{figure}

It has been shown by Halperin~\cite{Halperin:cikk} that the probability
current can be related to the derivative of the energy levels
with respect to the angular momentum quantum number:
\begin{equation}
I_{m,n} = \int_0^\infty\, {\bf j}\,  {\bf \hat{e}}_\varphi \, dr
= \frac{1}{h}\, \frac{\partial E_{m,n}}{\partial m}.
\label{Halperin_current:def}
\end{equation}
In Fig.~\ref{total_aram:fig} the integral of the current densities given
in Eq.~(\ref{Halperin_current:def}) is plotted as functions of $m$
for $n=0, 1, 2$. One can see that for $m\ll 0$ the total current
is zero (these are the orbits of type $B_1$ in our classification)
and that for $m\gg 2s$ it tends to a constant value (type $B_2$).
The current is negative for states corresponding to
orbits of type $A_1$, while it increases monotonically for states
corresponding to orbits of types $A_2 - A_4$.
This is in accordance with the right hand side of
Eq.~(\ref{Halperin_current:def}) using  the energy dispersion (the $m$
dependence of the energy levels) plotted
in Fig.~\ref{exact_levels-BS:fig}.

\section{Conclusions}\label{conclusion:sec}

In this paper we investigated the energy spectrum of the
circular magnetic anti-dot systems obtained from exact quantum and
semiclassical calculations.
In the proper dimensionless variables the only relevant parameter of the
system is the missing flux quantum.
The system is separable, and
in the quantum calculation the energy levels are the solutions of the
secular equation derived from the matching conditions of the
wave functions inside and outside the dot.
In the semiclassical treatment we presented two different methods.
On the one hand, the density of states was calculated using 
the Berry-Tabor formula.
On the other hand, the energy levels obtained 
from the Bohr-Sommerfeld quantization rules.
The main difference between the two methods is that in the first case
one needs to characterize the possible periodic orbits in real 
space, while in the latter, the motion of the particle is on a torus 
in the space of the action variables.
In our numerical results we compared the quantum energy spectrum and that
obtained in the semiclassical approach.
We showed that the energy levels of the magnetic anti-dot systems obtained
from the two semiclassical methods were in good agreement
with the numerically exact quantum results for weak magnetic fields.
However, by increasing the magnetic field,
a slight deviation between the exact and the semiclassically
approximated energy levels can be observed.
We argued that the reason for this discrepancy may be traced back 
to the fact that the Maslov index should be magnetic field dependent.
A thorough investigation of this problem might be an interesting future
work.

A classification of the energy spectrum for arbitrary magnetic
fields was presented in terms of the classical orbits
defined by their cyclotron radius and guiding center.
Such identifications are based on the explicit relations between these
classical parameters of the orbits and the quantum states.
The correspondence between the quantum states and the classical
trajectories can be made transparent by drawing a phase diagram 
with regions corresponding to six different types of orbits in the space
of energy and angular momentum quantum number.

Finally, we calculated the current flow patterns for eigenstates that
correspond to orbits with trajectories penetrating into the field-free
region. 
The related classical trajectories were also shown for the sake
of comparisson. From these results one can see the close correspondence
between the structure of the trajectories
and  the distribution of the current densities obtained from the quantum
calculations.

From the energy spectrum of the magnetic anti-dot systems one can
determine the free energy. 
The good agreement between the semiclassical and quantum
treatment of the system allows us to use semiclassical methods 
in the weak field limit for calculating the energy spectrum.
Therefore, the semiclassical approach provides a useful starting point
for successive studies of thermodynamic properties, such as magnetization.
Moreover, the semiclassical approximation can be an
effective tool for investigating arbitrarily shaped magnetic anti-dots
(which would be a very difficult task in the quantum case)
or systems with more complicated magnetic field profiles.

\acknowledgments

We would like to thank B. Kramer, A. Nogaret and A. Pir\'oth
for useful discussions.
This work is supported in part by the
European Community's Human Potential Programme under Contract No.
HPRN-CT-2000-00144, Nanoscale Dynamics, the Hungarian-British
Intergovernmental Agreement on Cooperation in Education, Culture,
Science and Technology, and the Hungarian  Science Foundation OTKA
TO34832 and FO47203.

\appendix

\section{The Berry-Tabor formula }
\label{Berry_Tabor:app}

The quantized energies can be recovered if we express the Hamiltonian
in terms of $I_i$
\begin{eqnarray}
& &E(n_1,n_2,...,n_d)=H(I_1,I_2,...,I_d)=\nonumber \\
& 
&H\left(\hbar\left(n_1+\nu_1/4\right),\hbar\left(n_2+\nu_2/4\right),...,\hbar\left(n_d+\nu_d/4\right)\right).\nonumber 
\\ & &
\end{eqnarray}
The semiclassical density of states is the density of these energies:
\begin{equation}
d(E)=\sum_{n_1,n_2,...,n_d=0}^{\infty}\delta\left(E-E(n_1,n_2,...,n_d)\right).
\end{equation}
The density of states can be rewritten via the Poisson resummation technique
\begin{eqnarray}
d(E)&=&\int d^dI\delta(E-H(I_1,I_2,...,I_d))\times\nonumber \\
& &\prod_{i=1}^{d}
\sum_{n_i=-\infty}^{+\infty}
\delta(I_i-\hbar(n_i+\nu_i/4))= \nonumber \\
& &
\sum_{m_1,m_2,...,m_d=-\infty}^{\infty}\int \frac{d^dIdt}{2\pi\hbar^{d+1}}
\times \nonumber \\ & &
e^{\frac{i}{\hbar}(t(E-H(I_1,...,I_d))+2\pi\sum_i m_i
(I_i-\hbar\nu_i/4))}. \label{altal}
\end{eqnarray}
Here, we used the Fourier expansion of the delta spike train. The term 
$m_i=0\;(i=1,2,...,d)$ can be evaluated directly and yields the 
non-oscillatory average density of states. Other terms can be
evaluated by the saddle point method, when $\hbar \rightarrow 0$. The
saddlepoint conditions select the periodic orbits of the system,
and the result of the integration is
\begin{eqnarray}
d(E)&=&d_0(E)+
\sum_{p}\sum_{j=1}^{+\infty}\frac{(2\pi)^{(d-1)/2}}{2^{
(\chi_p-1)}\hbar^{(d+1)/2}}\times \nonumber  \\ & &
\frac{\cos\left({jS_p(E)/\hbar}-\frac{\pi}{2}j\nu_p+\frac{\pi}{4}(d-1)
\right)}{\sqrt{(jT_p)^{d-1}(-\det {\bf D}_p)}}.
\label{density}
\end{eqnarray}
Here $p$ is the index of the primitive periodic orbits, $j$ is the number
of repetitions, $S_p$ is the classical action along the orbit,
$T_p$ is the time period of the orbit, and $\nu_p$ is the Maslov index.
The quantity $\chi_p$ is the number of action variables of the periodic
orbit whose saddle point value is zero $(I_k=0)$, since in this case
the Gaussian saddle point integral is only one-sided, and
its contribution is $1/2$ of the full Gaussian integral.
The matrix ${\bf D}_p$ is related to the second derivative matrix
\begin{equation}
\det {\bf D} =\det
\left(\begin{array}{cc}
\frac{\partial^2 H(I_1,...,I_d)}{\partial I_i\partial I_j}& \frac{\partial 
H(I_1,...,I_d)}{\partial I_i}\\
\frac{\partial H(I_1,...,I_d)}{\partial I_j}& 0\\
\end{array}\right).
\end{equation}
Equation ($\ref{density}$) is the generic form of the semiclassical
density of states in terms of periodic orbits, known as the Berry-Tabor
formula \cite{Berry-Tabor:cikk}.

In two dimensions, very often the Hamiltonian cannot be expressed with the
action variables
explicitly, only the implicit function
\begin{equation}
I_2=g(I_1,H),\label{gI}
\end{equation}
is available. In this case it is more useful to express the quantities
in the Berry-Tabor trace formula in terms of the derivatives of $g$.
Taking the partial derivative of (\ref{gI}) with respect to
$I_1$ yields
\begin{equation}
0=\frac{\partial g(I_1,H)}{\partial I_1}+\frac{\partial g(I_1,H)}{\partial 
H}
\frac{\partial H(I_1,I_2)}{\partial I_1},\label{gI1}
\end{equation}
while the partial derivative of (\ref{gI}) with respect to $I_2$ gives
\begin{equation}
1=\frac{\partial g(I_1,H)}{\partial H}
\frac{\partial H(I_1,I_2)}{\partial I_2}.\label{gI2}
\end{equation}
The frequencies can be expressed from these equations as
\begin{eqnarray}
\omega_1&=&\frac{\partial H(I_1,I_2)}{\partial I_1}=
-\frac{\frac{\partial g(I_1,H)}{\partial I_1}}{\frac{\partial 
g(I_1,H)}{\partial
H}},\\
\omega_2&=&\frac{\partial H(I_1,I_2)}{\partial I_2}=
\frac{1}{\frac{\partial g(I_1,H)}{\partial H}}.
\end{eqnarray}
Periodic orbits are recovered from
$\omega_1=\frac{2\pi n_1}{T}$ and $\omega_2=\frac{2\pi n_2}{T}$.
The action $I_1$ for a periodic orbit at energy $E$ can be obtained by 
solving
equation
\begin{equation}
\frac{\omega_1}{\omega_2}=\frac{n_1}{n_2}=\frac{n_{1,p}}{n_{2,p}}=-\frac{\partial 
g(I_1,E)}{\partial I_1},
\label{orbit}
\end{equation}
where we introduced $n_1=jn_{1,p}$ and $n_2=jn_{2,p}$
corresponding to the primitive orbit.
Then the period can be expressed simply as
\begin{equation}
T=2\pi n_{2,p}\frac{\partial g(I_1,H)}{\partial H}.
\end{equation}

The main determinant to be calculated reads
\begin{eqnarray}
\det {\bf D} &=&
\left|\begin{array}{ccc}
\frac{\partial^2 H(I_1,I_2)}{\partial I_1^2}&\frac{\partial^2 
H(I_1,I_2)}{\partial I_1\partial I_2}& \frac{\partial H(I_1,I_2)}{\partial 
I_1}\\
\frac{\partial^2 H(I_1,I_2)}{\partial I_1\partial I_2}&\frac{\partial^2 
H(I_1,I_2)}{\partial I_2^2}&\frac{\partial H(I_1,I_2)}{\partial I_2}\\
\frac{\partial H(I_1,I_2)}{\partial I_1}&\frac{\partial H(I_1,I_2)}{\partial 
I_2}& 0\\
\end{array}\right|
=\nonumber \\ &=& \left(-\frac{\partial^2 H}{\partial 
I_1^2}\left(\frac{\partial H}{\partial I_2}\right)^2
+2\frac{\partial^2 H}{\partial I_1\partial I_2}\frac{\partial H}{\partial 
I_1}\frac{\partial H}{\partial I_2}\right. \nonumber \\ & &
\left.
-\frac{\partial^2 H}{\partial I_2^2}\left(\frac{\partial H}{\partial 
I_1}\right)^2
\right).
\end{eqnarray}
Now, the second derivatives of $H$ can be expressed with the second
derivatives of $g$ by taking further partial derivatives of (\ref{gI1})
and (\ref{gI2}) with respect to $I_1$ and $I_2$.
Then we can express the second derivatives as
\begin{eqnarray}
\frac{\partial^2 H}{\partial I_1^2}&=&\frac{1}{\left(\frac{\partial 
g}{\partial H}\right)^3}\left(2\frac{\partial^2 g}{\partial H\partial I_1}
\frac{\partial g}{\partial I_1}\frac{\partial g}{\partial H}
-\frac{\partial^2 g}{\partial H^2}\left(\frac{\partial g}{\partial I_1}
\right)^2 \right.\nonumber \\ & &
\left.-\frac{\partial^2 g}{\partial I_1^2}\left(\frac{\partial g}{\partial 
H}\right)^2
\right),\\
\frac{\partial^2 H}{\partial I_1\partial 
I_2}&=&\frac{1}{\left(\frac{\partial g}{\partial H}\right)^3}\left(
\frac{\partial^2 g}{\partial H^2}\frac{\partial g}{\partial I_1}
-\frac{\partial^2 g}{\partial I_1\partial H}\frac{\partial g}{\partial H}
\right),\\
\frac{\partial^2 H}{\partial I_2^2}&=&-\frac{1}{\left(\frac{\partial 
g}{\partial H}\right)^3}\frac{\partial^2 g}{\partial H^2}.
\end{eqnarray}
Using these expressions, the determinant becomes
\begin{equation}
\det {\bf D}= -\frac{1}{\left(\frac{\partial g}{\partial H}\right)^3}
\frac{\partial^2 g}{\partial I_1^2}=-\frac{(2\pi 
n_{2,p})^3}{T^3}\frac{\partial^2 g}{\partial I_1^2}.
\label{det2}
\end{equation}

The density of states in two dimensions is then
\begin{eqnarray}
& &d(E)=d_0(E)+\nonumber \\ & &\sum_{p}\sum_{j=1}^{+\infty}
\frac{\cos\left({jS_p(E)/\hbar}-\frac{\pi}{2}j\nu_p+\frac{\pi}{4}(d-1)
\right)}{ 2^{\chi_p}\pi(\hbar)^{3/2}\sqrt{j(n_{2,p})^3
\frac{\partial^2 g}{\partial I_1^2}/T_p^2}}.\label{foalap}
\end{eqnarray}

\section{Derivation of the cyclotron radius and the guiding center}
\label{guiding:app}

The cyclotron radius can be determined from the energy $E$ of the
particle. The energy is conserved, and obviously 
$E = \frac{1}{2}\, M \omega_c^2 \, \varrho^2$, thus
\begin{equation}
\frac{\varrho}{l} = \sqrt{\varepsilon}.
\label{cycl-rad_app:eq}
\end{equation}

The guiding center may be calculated as follows.
As we have seen, the conjugate momentum given by Eq.~(\ref{pfi_phi_dot:eq})
is a constant of motion, therefore, e.g., for $r>R$
the right hand side of (\ref{pfi_phi_dot:eq})
should also be a constant at any point of the orbit.
At first apply this equation for points $P$ and $Q$,
which are the points closest to and farthest from the origin
(the center of the circle of radius $R$) of an orbit lying outside the
anti-dot.
These are special points of the orbits for which
the right hand side of Eq.~(\ref{pfi_phi_dot:eq}) has a simpler form.
Then the distances of points $P$ and $Q$
from the origin are
$r_P = c-\varrho$ and $r_Q = c+\varrho$ (we assume that point $Q$ is
farther from the origin).
From a simple geometrical argument one finds that the
angular velocity at points $P$ and $Q$ satisfies the
following equations
\begin{eqnarray}
r_P \, \dot{\varphi}_P &=& \varrho \, \omega_c \,  {\rm sgn}(eB),
\label{r_P_app:eq} \\
r_Q \, \dot{\varphi}_Q &=& -\varrho \, \omega_c \,  {\rm sgn}(eB).
\label{r_Q_app:eq}
\end{eqnarray}
Substituting, for example, $r_P = c-\varrho$ and $\dot{\varphi_P}$ from
(\ref{r_P_app:eq}) into Eq.~(\ref{pfi_phi_dot:eq}),
and using (\ref{cycl-rad_app:eq}), we find
\begin{equation}
\frac{c}{l} = \sqrt{\varepsilon +2 \, m_{\rm eff}}.
\end{equation}
The same results can be obtained by using (\ref{r_Q_app:eq}) for point $Q$.
If the oribit encompasses the anti-dot then the right hand side of
(\ref{r_P_app:eq}) should be multiplied by a factor of $-1$.
The case of orbits with trajectories penetrating into
the anti-dot can be treated similarly.
However, the expressions for the cyclotron radius and the guiding
center are the same as above  for all cases.

\section{Contribution of the cyclotron orbits to the semiclassical
density of states }  \label{Landau:app}

In the case of the cyclotron orbits, the integral in $I_{\varphi}$
in Eq. (\ref{altal}) has to be calculated directly rather than using
the saddlepoint
method. As $I_{\varphi}$ is constant, the integrand does not depend on
the integration variable and therefore the integral is equal to
the measure of the interval of the possible $I_{\varphi}$s.
Without loss of generality, we take $\rm{sgn}(eB)=-1$ in this section.

\subsection{Cyclotron orbits of type $B_1$}
Cyclotron orbits which do not encompass the anti-dot (type $B_1$)
are possible at any value of $\rho$ and any negative angular momentum 
$p_\varphi$ (see Table \ref{table:orbits_A-B-conds}).
At point $P$  of these orbits (points $P$ and $Q$  of
a cyclotron orbit are
defined in Appendix \ref{guiding:app}), from Eqs.~(\ref{angular_action:eq})
and (\ref{pfi_phi_dot:eq}) we obtain
\begin{eqnarray}
I_{\varphi} = p_{\varphi}=Mr_P^2\dot{\varphi}_P+eB\frac{r_P^2-R^2}{2}.
\label{I_phi_B_1;eq}
\end{eqnarray}
This is minimal when $r_P=R$ (and the cyclotron orbit touches the
boundary of the anti-dot), and is maximal
when  the orbit is placed as far as possible from the anti-dot.
By denoting the radius of the system with ${\cal L}$ (in units of $l$)
the intergation in (\ref{altal}) with respect 
to $I_{\varphi}$ yields a factor 
$\Delta  I_{\varphi} =  I_\varphi(R) - I_\varphi({\cal L})$.
Using  (\ref{I_phi_B_1;eq}) and (\ref{r_P_app:eq}), one finds
\begin{eqnarray}
\frac{\Delta I_\varphi}{\hbar}  =
{\cal L}\sqrt{\varepsilon}-\sqrt{2s\varepsilon}
+\frac{1}{2}({\cal L}^2-2s) .
\end{eqnarray}
The $I_r$- and $t$-integrals can be
evaluated with the saddle point method, just as 
in the case of a one-dimensional system.
The determinant of the second derivative matrix is
\begin{equation}
{\det}\begin{bf} D \end{bf}=\det\begin{pmatrix}-\frac{\partial^2H}{{\partial
}I_r^2}T & -\frac{{\partial}H}{
{\partial}I_r} \cr -\frac{{\partial}H}{{\partial}I_r} & 0
\end{pmatrix}=-\left(
\frac{{\partial}H}{{\partial}I_r}\right)^2.
\end{equation}
From Eqs.~(\ref{gI2}) and (\ref{I_r_B-type:eq}), we find
\begin{eqnarray}
\frac{{\partial}H}{{\partial}I_r}
&=& \frac{1}{\frac{{\partial}g}{{\partial}E}}=\omega_c , \\[1ex]
\det\begin{bf} D \end{bf} &=& - \omega_c^2.
\end{eqnarray}
Thus, the total amplitude of these orbits in the periodic orbit
sum is
\begin{eqnarray}
A_c^- &\equiv & \frac{\Delta I_\varphi}{\hbar} \, \frac{1}{\hbar \omega_c}
\nonumber \\
&=& \frac{1}{ \hbar \omega_c }
\left[{\cal L}\sqrt{\varepsilon}-\sqrt{2s\varepsilon}
+\frac{1}{2}({\cal L}^2-2s) \right].
\label{eq:A_c_pl}
\end{eqnarray}
The action can be calculated from Eq.~(\ref{I_r_B-type:eq}), and for
$p_\varphi/\hbar \equiv m  <0$ we have
\begin{eqnarray}
S= 2 \pi I_r =  \hbar \pi \varepsilon,
\label{action_B_1:eq}
\end{eqnarray}
and their Maslov index is $\mu=2$, therefore the contribution 
to the semiclassical level density from these orbits reads
\begin{eqnarray}
d_c^-(E)=A_c^-\sum_{j=1}^{\infty}
\cos\left(j\varepsilon\pi-j\pi\right).
\label{eq:cyc_contr_alap}
\end{eqnarray}
The sum has Dirac delta peaks at $\varepsilon = 2n+1$, where
$n=0, 1, 2, \ldots$ and $m<0$.
These are the familiar Landau levels of an electron for $m<0$.

\subsection{Cycltoron orbits of type $B_2$ }
For cyclotron orbits encompassing the anti-dot (type $B_2$)
the angular momentum satisfies the condition
$p_\varphi /\hbar = m> 2s $ (see Table \ref{table:orbits_A-B-conds}).
At point $Q$ of these orbits, using
(\ref{angular_action:eq}) and (\ref{pfi_phi_dot:eq}), we can write
\begin{eqnarray}
I_{\varphi} = p_{\varphi}=Mr_Q^2\dot{\varphi_Q}+eB\frac{r_Q^2-R^2}{2}.
\label{I_phi_B_2;eq}
\end{eqnarray}
The minimum and the maximum of $r_Q$ are $\varrho$ and $2\varrho -R$,
respectively. Between these values, $I_{\varphi} $, as a function of
$r_Q$, is monotonic, thus the integration in (\ref{altal}) over
$I_{\varphi}$ gives
$\Delta  I_{\varphi} = I_\varphi(\varrho) - I_\varphi(2\varrho -R)$.
Using (\ref{I_phi_B_2;eq}) and  (\ref{r_Q_app:eq}), we have
\begin{eqnarray}
\frac{\Delta I_\varphi}{\hbar} =
\frac{\varepsilon}{2} + s - \sqrt{2s\varepsilon}.
\end{eqnarray}
Similarly to (\ref{eq:A_c_pl}), the amplitude of the orbits becomes
\begin{equation}
A_c^+ \equiv  \frac{\Delta I_\varphi}{\hbar} \,
\frac{1}{\hbar \omega_c}
= \frac{\frac{\varepsilon}{2} + s - \sqrt{2s\varepsilon}}
{\hbar \omega_c }.
\end{equation}
Using (\ref{I_r_B-type:eq}), the action for
$p_\varphi/\hbar \equiv m  > 2s $ is
\begin{eqnarray}
S= 2 \pi I_r =  \hbar \pi \left(\varepsilon + 2s -2m \right),
\end{eqnarray}
Finally, the contribution to the periodic orbit sum of these orbits is
\begin{eqnarray}
d_c^+(E)&=&  \!\! A_c^+\sum_{j=1}^{\infty}
\cos\left[\pi j(\varepsilon+2s-2m)-j\pi \right].
\end{eqnarray}
The sum has Dirac delta peaks at $\varepsilon = 2(m-s)+2n+1$, where
$m$ and $n$ are non-negative integers, and $m>2s$.
These are again the familiar Landau levels of an electron for $m>2s$.

\end{document}